\documentclass[aps,prev,twocolumn,preprintnumbers,floatfix,nofootinbib]{revtex4-1}

\usepackage[T1]{fontenc} 
\usepackage[utf8]{inputenc}
\usepackage{physics}

\usepackage[dvipdfm]{graphicx}
\usepackage{mathrsfs}
\usepackage{amssymb}
\usepackage{verbatim}
\usepackage{color}
\usepackage[dvipsnames]{xcolor}
\usepackage{multirow}
\usepackage{amsmath}
\usepackage{slashed} 
\usepackage[normalem]{ulem}
\usepackage{sidecap}
\usepackage{hyperref}
\usepackage{cancel}
\usepackage{enumerate}
\hypersetup{pdfstartview=FitV,colorlinks=true,linkcolor=blue,citecolor=red,filecolor=black,urlcolor=blue}

\newcommand{\aend}{a_{\rm end}}
\newcommand{\arh}{a_{\rm RH}}
\newcommand{\am}{a_{\rm m}}
\newcommand{\af}{a_{\rm F}}
\newcommand{\Trh}{T_{\rm RH}}
\newcommand{\rhorh}{\rho_{\rm RH}}
\newcommand{\rhoe}{\rho_{\rm end}}
\newcommand{\phie}{\phi_{\rm end}}
\newcommand{\rhom}{\rho_{\rm m}}
\newcommand{\rhor}{\rho_{\rm R}}
\newcommand{\beq}{\begin{equation}}
\newcommand{\eeq}{\end{equation}}
\newcommand{\bea}{\begin{eqnarray}}
\newcommand{\eea}{\end{eqnarray}}

\def\trh{T_{\rm RH}}

\usepackage[parfill]{parskip}
\begin{document}\sloppy

 \preprint{UMN--TH--4325/24, FTPI--MINN--24/06}

\vspace*{1mm}

\title{Bare mass effects on the reheating process after inflation }
\author{Simon Clery$^{a}$}
\email{simon.clery@ijclab.in2p3.fr}
\author{Marcos~A.~G.~Garcia$^{b}$}
\email{marcos.garcia@fisica.unam.mx}
\author{Yann Mambrini$^{a}$}
\email{yann.mambrini@ijclab.in2p3.fr}
\author{Keith A. Olive$^{c}$}
\email{olive@umn.edu}
\vspace{0.5cm}
\affiliation{${}^a $  Universit\'e Paris-Saclay, CNRS/IN2P3, IJCLab, 91405 Orsay, France}
\affiliation{${}^b$Departamento de F\'isica Te\'orica, Instituto de F\'isica, Universidad Nacional Aut\'onoma de M\'exico, Ciudad de M\'exico C.P. 04510, Mexico}
\affiliation{${}^c$William I. Fine Theoretical Physics Institute, School of
 Physics and Astronomy, University of Minnesota, Minneapolis, MN 55455,
 USA}

\date{\today}

\begin{abstract} 
We consider the effects of a bare mass term for the inflaton, when the inflationary potential takes the form $V(\phi)= \lambda \phi^k$ about its minimum with $k \ge 4$. We concentrate on $k=4$, but discuss general cases as well. Further, we assume $\lambda \phi_{\rm end}^2 \gg m_\phi^2$, where $\phie$ is the inflaton field value when the inflationary expansion ends. 
We show that the presence of a mass term (which may be present due to radiative corrections or supersymmetry breaking) can significantly alter the reheating process, as the equation of state of the inflaton condensate changes from $w_\phi=\frac{1}{3}$ to $w_\phi=0$
when $\lambda \phi^2$ drops below $m_\phi^2$. We show that for a mass $m_\phi \gtrsim \trh/250$, the mass term will dominate at reheating. We compute the effects on the reheating temperature for cases where reheating is due to inflaton decay (to fermions, scalars, or vectors) or to inflaton scattering (to scalars or vectors). For scattering to scalars and in the absence of a decay, we derive a strong upper limit to the inflaton bare mass $m_\phi < 350~{\rm MeV} (\trh/10^{10}~{\rm GeV})^{3/5}$, as there is always a residual inflaton background which acts as cold dark matter. 
We also consider the effect of the bare mass term on the fragmentation of the inflaton condensate. 
\end{abstract}

\maketitle

\setcounter{equation}{0}

\section{Introduction}

The hypothesis 
of a violent inflationary phase during the first moments of the 
Universe makes it possible to address several cosmological 
issues, ranging from the flatness of the Universe to the 
horizon or entropy problem~\cite{reviews}. However, a complete
inflationary model requires above all a mechanism for a graceful exit.
Indeed, the prolonged period of exponential expansion must 
end with a sufficiently efficient transfer of the 
oscillation modes of the 
inflaton condensate $\phi$ to a thermal bath \cite{dg,nos}, i.e. reheating,  that 
ensures a temperature $\gtrsim 2$ MeV to allow for standard big bang nucleosynthesis. 
Moreover, the density fluctuation spectrum produced during 
inflation should agree with observations of the CMB anisotropy spectrum~\cite{Planck}, 
which in turn constrains the parameters of the inflaton potential  $V(\phi)$. 

The process of transferring the energy stored in inflaton oscillations to Standard Model 
particles is not instantaneous~\cite{Giudice:2000ex,Mukaida:2015ria,GMOP,Bernal:2020gzm}. Rather, in many models, an oscillating inflaton condensate decays or scatters 
progressively producing a bath of relativistic particles. 
The efficiency of the reheating process depends on the 
rate of the energy transfer as well as on the shape
of the inflaton potential, $V(\phi)$, about its minimum \cite{GKMO1,GKMO2}.
Even if the exact shape of the 
potential at the end of inflation is unknown 
it can often be approximated about its minimum by a polynomial function of $\phi$.

In many models of inflation, the inflaton potential can be approximated about its minimum by a   
quadratic term, $V(\phi)=\frac{1}{2}m_\phi^2\phi^2$. The Starobinsky model \cite{Staro} is one example. In this case,  only one Fourier mode of the inflaton oscillation contributes to the reheating process. 
The energy density in radiation, $\rho_R$, 
grows rapidly at first, and redshifts 
as $\rhor\propto a^{-\frac{3}{2}}$ 
where $a$ is the cosmological scale factor, as decays continue to add to the radiation bath. 
Because $\rho_\phi \propto a^{-3}$, eventually, the radiation bath comes to dominate the total 
energy density, at which time we can define a reheating temperature. This occurs when the cosmological 
scale factor, $\arh$ satisfies
$\rho_{\rm R}(\arh) = \rho_\phi(\arh)$.  
This occurs
(up to a numerical factor) when
$H(\arh)\simeq \Gamma_\phi$, or $\Trh\simeq \sqrt{\Gamma_\phi M_P}$, where $H$ is the Hubble parameter,  $\Gamma_\phi$ is the width of the 
inflaton condensate, and $M_P= 1/\sqrt{8 \pi G_N}\simeq 2.4\times 10^{18}$ GeV is the reduced Planck mass. 

For a potential whose expansion about its minimum is
$V(\phi)=\lambda \phi^k$, with $k \ge 4$, the exercise is more subtle,
and requires a more involved analysis~\cite{GKMO1,GKMO2}.
The reheating process will in general depend on the spin of the final state particles in either inflaton decays or scatterings. In fact, in some cases reheating does not occur. 
For example, for $k=4$, the evolution of $\rho_\phi \propto a^{-4}$
is the same as the evolution of $\rhor \propto a^{-4}$ for inflaton decays or scatterings to vector bosons \cite{gkkmov}, precluding
the condition $\rho_\phi(\arh)=\rho_R(\arh)$ to occur.
However,  we cannot exclude the presence of a bare mass term $\frac{1}{2}m_\phi^2\phi^2$, which may be 
subdominant at the end of inflation, and during the early phases of the oscillations, but which becomes dominant when 
$\phi$ has redshifted down to a point $\am$ defined by 
$\lambda \phi^4(\am) = \frac{1}{2}m_\phi^2\phi^2 (\am)$. The presence of this term, 
even if it is small, would then modify the reheating 
mechanisms, making for example reheating by decays to vector bosons possible in the case $k=4$. 

Many models of inflation have potentials which, when expanded about their minimum, are described by a series of self interactions beyond their mass term. 
For example, the well studied Starobinsky potential 
\cite{Staro}, contains a full series of interaction terms. However, for $\phi < \phi_{\rm end}$, where $\phie = \phi(\aend)$ is the inflaton field value when the inflationary expansion ends (when $\ddot{a} = 0$),
terms which are higher order than the quadratic (mass term), become greatly suppressed and do not substantially affect the subsequent evolution of the inflaton condensate. In contrast, models such the so-called $\alpha$-attractor $T$-models of inflation \cite{Kallosh:2013hoa}, described by a potential of the form,
\begin{equation}
    V(\phi) \; = \;\lambda M_P^{4}\left|\sqrt{6} \tanh \left(\frac{\phi}{\sqrt{6} M_P}\right)\right|^{k} \, ,
\label{Vatt}
\end{equation}
contain only even interaction terms starting with $\lambda M_P^{4-k} \phi^k$ yielding a massless inflaton for $k \ge 4$. 

A bare mass term may be present at the tree level,
may be produced as a result of supersymmetry breaking in a supersymmetric model, or may be produced radiatively. Though we will treat the mass as a free parameter, we note that there are 1)
upper limits on the mass imposed by slow-roll parameters which determine the inflationary observables, $n_s$ and $r$; 2) in the absence of fine-tuning, there is a lower bound on the mass derived from 
loop corrections to the potential {\` a} la Coleman-Weinberg. 
Both of these limits will be discussed below. In any case,
the presence of a mass term seems unavoidable, at least 
at higher order, justifying a detailed analysis of its effect on the reheating process.

More specifically, the reheating phase in the $T$-models with $k\ge4$, as an example, is altered when a mass term is added to the potential in Eq.~\ref{Vatt}. As a result,  for $k=4$, the evolution of the energy density transitions from a radiation-dominated Universe ($V(\phi)\propto \phi^4$, $\rho_\phi\propto a^{-4}$) to a matter-dominated Universe ($V(\phi)\propto \phi^2$, $\rho_\phi\propto a^{-3}$).\footnote{More generally, the Universe transitions from an expansion with an equation of state, $w=P_\phi/\rho_\phi = (k-2)/(k+2)$ ($V(\phi)\propto \phi^k$, $\rho_\phi\propto a^{-6k/(k+2)}$) to a matter-dominated Universe. }
If the reheating process is sufficiently slow, the quadratic term can come to dominate the inflaton energy density and would result in higher reheating temperature than would have been achieved from the quartic term alone.
The presence of a bare mass term generalizes previous results \cite{GKMO1,GKMO2}. 

Furthermore, it was recently 
shown in \cite{GGMOPY,Garcia:2023eol} that the effects of the 
fragmentation of the inflaton condensate through its self interaction $\lambda \phi^k$, $k \geq 4$ 
could considerably affect the reheating 
process. It was noticed that in the case of reheating generated via fermion 
decay, fragmentation stopped the reheating 
process too early, leaving the Universe with a bath of massless (and thus stable) particles (inflatons). 
This would be in contradiction with CMB/BBN observations.
However, if the quadratic term $\frac{1}{2}m_\phi^2\phi^2$ were to dominate {\it before} 
the end of the fragmentation of the inflaton ($\am<\af$ where $\af$ is the scale factor when fragmentation is complete), the latter 
would stop, allowing the condensate to continue the reheating process safely
through its decay, $\Gamma_\phi\propto m_\phi$.  For example, for $k=4$, the conformal self-resonance responsible for the exponential growth of a narrow range of relativistic $\phi$ momentum modes is shut down as they become non-relativistic. The effective frequencies lose the oscillatory driving, and become incapable of fragmenting the inflaton condensate~\cite{GGMOPY,Garcia:2023eol}. This will be discussed in more detail below.

The paper is organized as follows: in Section \ref{Sec:am}, we describe the effect of the transition from a $\phi^4 \to \phi^2$ potential on the evolution of the inflaton condensate and its impact on the reheating temperature. In Section \ref{masslim}, we derive the upper limit to the inflaton mass from CMB observables and the bare mass expected from radiative corrections which in the absence of fine-tuning represents a lower limit to the mass. Then in Section \ref{Sec:coupling},
we derive the relations between the inflaton coupling to matter and the reheating temperature in view of the transition to a matter dominated expansion. These results are generalized to $k\ne4$ in Section \ref{sec:gen} and the consequences on the fragmentation of the inflaton condensate are discussed in Section \ref{sec:frag}. Our summary is found in Section \ref{sec:sum}.


\section{The transition, $\phi^4 \rightarrow \phi^2$}
\label{Sec:am}


We begin by supposing that the dominant contribution in a series expansion of the inflaton potential about its minimum is the quartic term and that at the end of inflation, this dominates over a quadratic mass term, so that
\beq
\lambda \phie^4 \gg \frac{1}{2}m_\phi^2 \phie^2\,.
\eeq
For $a > \aend$, the evolution of the energy density of $\phi$ is governed by
the Friedmann equation for $\rho_\phi$
\beq
\frac{d \rho_\phi}{dt}+ 3(1+w) H \rho_\phi\simeq 0 \, .
\label{Eq:frphi}
\eeq
Where $\rho_\phi=\langle V(\phi) \rangle=V(\phi_0)$, the mean being taken over the oscillation of $\phi$ and $\phi_0$ is the envelope of the oscillations. More precisely,

\beq
\phi(t) = \phi_0(t)\, \mathcal{P}(t)\,,
\eeq
with $\mathcal{P}(t)$ a quasiperiodic function encoding the (an)harmonicity of short-timescale oscillations in the potential.

For $k=4$, Eq.~(\ref{Eq:frphi}) gives
\beq
\rho_\phi = \rhoe\left(\frac{\aend}{a}\right)^4 \, ,
\eeq
where $\rhoe$ is the value of the density of energy of the inflaton at the end of inflation, when $\ddot{a}=0$. This condition is equivalent to $w=-1/3$ or $\dot{\phi}^2_{\rm end}=V(\phi_{\rm end})$. Hence, 
\beq
\rhoe=\frac32 V(\phie) \, ,
\label{rhoe}
\eeq
where for the $T$-models with potential given in Eq.~(\ref{Vatt}) we have \cite{GKMO2}, 
\beq
\phi_{\rm end} \;\simeq\;\sqrt{\frac{3}{8}}\, M_P \ln\left[ \frac{1}{2} + \frac{k}{3}\left(k+\sqrt{k^2+3}\right) \right]\,.
\eeq

The parameter $\lambda$ in Eq.~(\ref{Vatt}) is determined from the normalization of the
CMB anisotropies \cite{Planck}. 
The normalization of the potential for different values of $k$ can be approximated by~\cite{GKMO2}
\begin{equation}
    \label{eq:normlambda}
    \lambda \; \simeq \; \frac{18\pi^2 A_{S*}}{6^{k/2} N_*^2} \, ,
\end{equation}
where $N_*$ is the number of e-folds from horizon crossing to the end of inflation and $A_{S*} \simeq 2.1 \times 10^{-9}$ is the amplitude of the curvature power spectrum. For $N_* = 56$ e-folds we find $\lambda=3.3 \times 10^{-12}$, and $\rhoe^{\frac{1}{4}}=4.8\times 10^{15}$ GeV (when $k=4$).

As $\phi_0$ decreases, eventually the evolution of the condensate will be governed by the quadratic term. This occurs at $a=\am$ when
\beq
\frac{1}{2}m_\phi^2\phi_0^2(\am)=\lambda \phi_0^4(\am) \, .
\eeq
Using $\phi_0^4(a) = (\rhoe/\lambda) (\frac{\aend}{a})^4$ for $\aend < a < \am$ gives
\beq
\frac{\am}{\aend}=
 \left(\frac{4 \lambda \rhoe}{m_\phi^4}\right)^{1/4}
\simeq 9.1 \times 10^3 \left(\frac{10^{9}~\rm{GeV}}{m_\phi}\right)\, .
\label{Eq:am}
\eeq
In deriving (\ref{Eq:am}), we note that the envelope function $\phi_0$ is determined by the average energy density $\langle \rho_\phi \rangle = V(\phi_0)$. Thus unless reheating occurs rapidly,
the quadratic term will dominate the reheating process even if the quartic dominates after when oscillations begin. This will have 
huge consequences on the reheating temperature, as well as on the physics of fragmentation as we will see.



Indeed, if reheating occurs at $a = \arh > \am$, the process is affected by the bare mass term.
For $a>\am$, the equation of state changes from
$w=1/3$ (for $k=4$) to $w=0$ (for $k=2$) 
and the solution for $a \ll a_m$ to the Friedmann equation becomes
\beq
\rho_\phi= \frac12 \rho_\phi(\am) \left(\frac{\am} {a}\right)^3 = \rhoe \left(\frac{\aend} {\am}\right)^4 \left(\frac{\am} {a}\right)^3 \,.
\label{Eq:rhophibis}
\eeq
Furthermore,
\beq
\rhom \equiv  \rho_\phi(\am)= 2\rhoe\left(\frac{\aend}{\am}\right)^4=
\frac{ m_\phi^4}{2 \lambda}\,.
\label{Eq:rhom}
\eeq
Combining Eqs.~(\ref{Eq:am}) and (\ref{Eq:rhophibis})
we obtain
\beq
\left.\rho_\phi\right|_{a>\am}= \frac{m_\phi \rhoe^{\frac{3}{4}} }{\left(4\lambda\right)^{\frac{1}{4}}}\left(\frac{\aend}{a}\right)^3\,.
\eeq

This form for $\rho_\phi$ dominates the energy density until reheating defined by $\rho_\phi(\arh)=\rhor(\arh)$.  
Here, $\rhor$ is the energy density 
transferred to the thermal bath via the Boltzmann equation
\beq
\frac{d \rho_R}{dt} + 4 H \rho_R = (1+w)\Gamma_\phi \rho_\phi\,.
\label{Eq:radiation_density}
\eeq
From the above, we can determine the reheating temperature for a given mass, $m_\phi$ for which  the bare mass affects the reheating process, and therefore modifies the calculation of 
$\Trh$.  The condition $\am<\arh$ implies that $\rhom > \rho_\phi(\arh)$ and thus
the condition for the quadratic part to dominate the reheating process is given by 
\beq
\rhom\gtrsim\rhorh~~\Rightarrow ~~\rhorh\lesssim\frac{m_\phi^4}{2 \lambda}\,,
\eeq
from Eq.(\ref{Eq:rhom}).

Defining $\rhorh=\alpha\Trh^4$ with $\alpha=\frac{g_{\rm RH} \pi^2}{30}$
for $g_{\rm RH}$ relativistic degrees of freedom at $\arh$, 
we obtain
\beq
\trh \lesssim \frac{m_\phi}{(2 \alpha \lambda)^\frac{1}{4}}\simeq 250~m_\phi\,,
\label{Eq:limittrh}
\eeq
which means that if the energy transfer between the condensate and the 
thermal bath is slow and the reheating temperature $\Trh$ lower than the limit obtained in the equation (\ref{Eq:limittrh}), we 
must take into account the quadratic term to determine $\trh$ when $\arh > \am$.

\section{Limits on the inflaton bare mass
}
\label{masslim}

As noted earlier, the CMB observables impose an upper limit to $m_\phi$ and in the absence of any fine-tuning, couplings of the inflaton to Standard Model fields (necessary for reheating), provide a lower bound to $m_\phi$ from radiative corrections to the potential. 

Planck \cite{Planck} has determined with relatively high precision, the value for the tilt of the CMB anisotropy spectrum, $n_s = 0.9649 \pm 0.0042$ (68\% CL). In addition, the tensor-to-scalar ratio, $r < 0.036$ is constrained by BICEP/Keck observations \cite{BICEP2021,Tristram:2021tvh}.
To translate these limits to an upper limit on $m_\phi$, we use the $T$-model in Eq.~(\ref{Vatt}) as an example. 

Recall that the conventional slow-roll parameters
for a single-field inflationary model are given by
\begin{equation}\label{eq:epseta}
\epsilon \equiv \frac{1}{2} M_{P}^{2}\left(\frac{V^{\prime}}{V}\right)^{2} \, , \qquad \eta \equiv M_{P}^{2}\left(\frac{V^{\prime \prime}}{V}\right) \, ,
\end{equation}
where the prime denotes a derivative with respect to the inflaton field, $\phi$. The number of $e$-folds can be computed using
\begin{equation}
\label{eq:efolds}
N_{*} \simeq \frac{1}{M_{P}^{2}} \int_{\phi_{\mathrm{end}}}^{\phi_{*}} \frac{V(\phi)}{V^{\prime}(\phi)} d \phi \simeq \int_{\phi_{\mathrm{end}}}^{\phi_{*}} \frac{1}{\sqrt{2 \epsilon}} \frac{d \phi}{M_{P}} \, ,
\end{equation}
where $\phi_*$ corresponds to the horizon exit scale $k_* \; = \; 0.05 \, \rm{Mpc}^{-1}$ used in the {\it Planck} analysis.
The scalar tilt and tensor-to-scalar ratio can be expressed in terms of the slow roll parameters as
\begin{align}
    \label{eq:spectrtilt}
    n_{s} \; &\simeq \; 1-6 \epsilon_{*}+2 \eta_{*} \, , \\
    \label{eq:sclrtotens}
    r \; &\simeq \; 16 \epsilon_{*} \, .
\end{align}
In a more precise model determination of $N_*$, and $n_s$, there is some dependence on the reheating temperature and equation of state \cite{LiddleLeach,Martin:2010kz}. The computation is based on the self-consistent solution of the relation between $N_*$ and its corresponding pivot scale $k_*$,
\begin{align}\notag
N_{*} \; = \; &\ln \left[\frac{1}{\sqrt{3}}\left(\frac{\pi^{2}}{30}\right)^{1 / 4}\left(\frac{43}{11}\right)^{1 / 3} \frac{T_{0}}{H_{0}}\right]-\ln \left(\frac{k_{*}}{a_{0} H_{0}}\right) \\ \notag
& -\frac{1}{12} \ln g_{\rm RH}  +\frac{1}{4} \ln \left(\frac{V(\phi_{*})^{2}}{M_{P}^{4} \rho_{\mathrm{end}}}\right) \\ \label{eq:nstar}
& + \ln\left[\frac{a_{\rm end}}{a_{\rm RH}}\left(\frac{\rho_{\rm end}}{\rho_{\rm RH}}\right)^{1/4}\right]\,,
\end{align}
where the present Hubble parameter and photon temperature are given by $H_0 = 67.36 \, \rm{km \, s^{-1} \, Mpc^{-1}}$~\cite{Planck} and $T_0 = 2.7255 \, \rm{K}$~\cite{Fixsen:2009ug}.  For the $T$-Models dominated by a quadratic term, agreement with Planck/BICEP/Keck data requires $N_*$ between roughly 42 - 56 \cite{egnov}.

\begin{figure}[!t]
\centering
\includegraphics[width=\columnwidth]{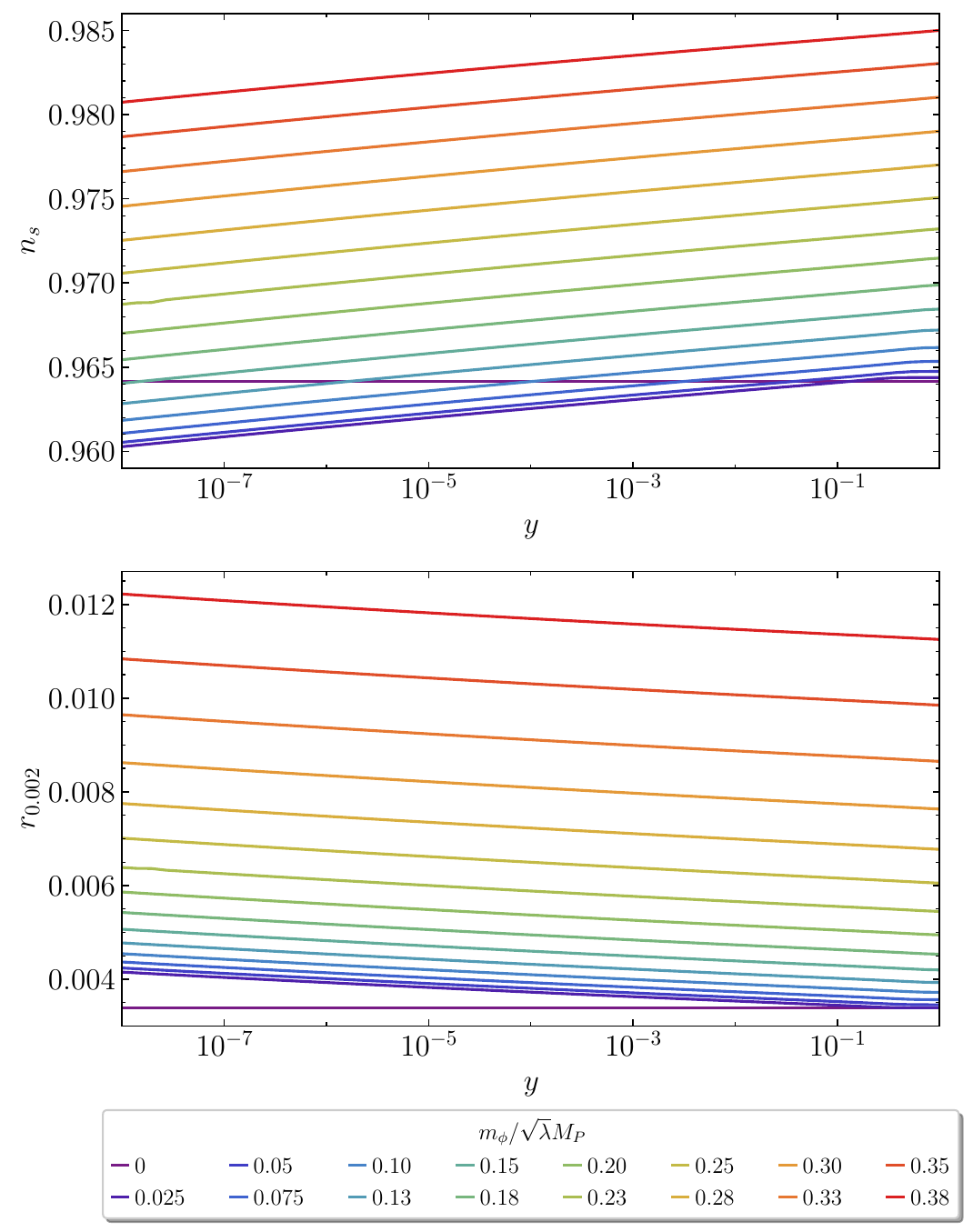}
\caption{\em \small Scalar tilt $n_s$ (top) and tensor-to-scalar ratio $r$ (bottom) as functions of the Yukawa coupling $y$ (\ref{eq:ydef}), for a selection of bare masses $m_{\phi}$ and $k=4$.
}
\label{Fig:nsry}
\end{figure}

\begin{figure}[!ht]
\centering
\includegraphics[width=\columnwidth]{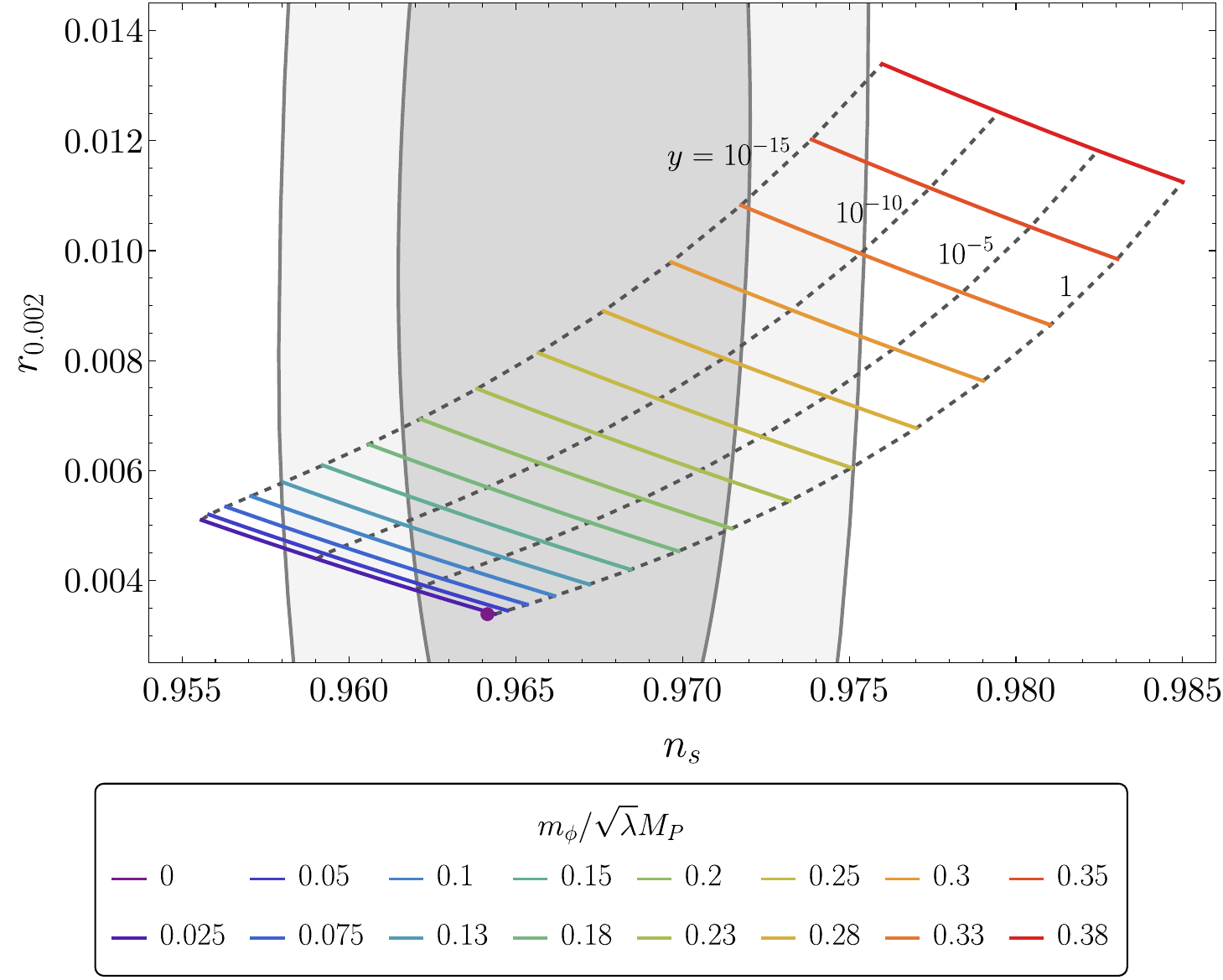}
\caption{\em \small Same as Fig.~\ref{Fig:nsry} shown in the $(n_s,r)$ plane. The gray (light gray) shaded regions correspond to the 68\% (95\%) C.L. Planck+BK18 regions~\cite{BICEP2021}.
}
\label{Fig:nsrp}
\end{figure}

In the absence of a mass, $m_\phi = 0$, $N_*\simeq 56$ with $\phi_*=6.96 M_P$ and $(n_s,r)=(0.964,0.0034)$,  independently of the efficiency of reheating~\cite{GKMO2,Garcia:2023eol}. Therefore, to set limits on a possible mass term for $k=4$, we set $N_* = 56$. For non-zero masses both $n_s$ and $r$ increase, but the limit on $m_\phi$ is determined mainly from $n_s$. Figs.~\ref{Fig:nsry} and \ref{Fig:nsrp} show the numerically computed CMB observables $n_s$ and $r$ for a variety of bare masses and inflaton-matter couplings. As is customary, the {\em Planck} $(k_*=0.05\,{\rm Mpc}^{-1})$ and WMAP $(k_*=0.002\,{\rm Mpc}^{-1})$ pivot scales are chosen for $n_s$ and $r$, respectively. For $m_{\phi}\neq 0$, the effective equation-of-state parameter evolves as $w= -1/3\rightarrow 1/3 \rightarrow 0\rightarrow 1/3$ from the end of inflation to the end of reheating. The top panel of Fig.~\ref{Fig:nsry} depicts the bare mass dependence of the scalar tilt, as a function of an inflaton-matter Yukawa coupling (see Eq.~(\ref{eq:ydef})). For $m_{\phi}=0$, $N_*\simeq 56$ for any $y$, leading to the purple horizontal line. For $m_{\phi}=0.025\sqrt{\lambda} M_P$, the smallest non-zero mass in the Figure, the resulting curve presents two regimes. At $y\gtrsim 10^{-1}$, $n_s$ is independent of $y$ since reheating is completed before matter domination, $a_{\rm RH}<a_{\rm m}$. However, for $y\lesssim 10^{-1}$, reheating is completed by the dissipation of the quadratic, harmonic oscillations of $\phi$. A dependence of $n_s$ on $y$ is induced, since now the last term of (\ref{eq:nstar}) is relevant for the determination of $N_*$. For smaller $y$ reheating is delayed, resulting in a smaller $N_*$ and as a consequence $n_s$. In the case of larger masses, the pure quartic regime is reduced, or outright lost, and the relation of $n_s$ and $y$ is determined by the duration of reheating in the matter dominated era, and the modification of the slow roll dynamics due to the presence of the large bare mass. Analogous conclusions can be drawn from the bottom panel of Fig.~\ref{Fig:nsry}. In this case the addition of the bare mass increases the value of the tensor-to-scalar ratio, both from the modified inflation dynamics, and from the dependence on $y$ of the number of $e$-folds $N_*$.

 Fig.~\ref{Fig:nsrp} compares the corresponding $(n_s,r)$ curves against the {\em Planck}+BK18 constraints~\cite{BICEP2021}. Here the range of couplings spans reheating temperatures from $T_{\rm RH}\sim 2\times 10^{14}$~{\rm GeV} for $y= 1$, to $T_{\rm RH}\sim \mathcal{O}(10)\,{\rm MeV}$ for $y=10^{-15}$. We note that for the smallest bare masses high reheating temperatures are favored by the CMB data. On the other hand, for the largest masses considered, lower $T_{\rm RH}$ are preferred. At the nominal $N_*=56$, corresponding to $y\approx 1$ in the figure, we find that $m_\phi < 0.2 \sqrt{\lambda} M_P \simeq 8.8 \times 10^{11}$ GeV at 68\% CL with $(n_s,r)= (0.971,0.0050)$ and $m_\phi < 0.25 \sqrt{\lambda} M_P \simeq 1.1 \times 10^{12}$ GeV at 95\% CL with $(n_s,r)= (0.975,0.0061)$. Above 
these masses, the values of $n_s$ and $r$ rise very quickly and agreement with data is lost. Applying this limit on $m_\phi$ in Eq.~(\ref{Eq:limittrh}) gives $\trh \lesssim 2.8 \times 10^{14}$ GeV. In other words, for larger reheating temperatures, the energy transfer is sufficiently efficient to avoid any interference of a possible quadratic interaction without violating the CMB data. Allowing for the full range in coupling $y$ or equivalently $\trh$ and expanding the range in $N_*$, 
we see from Fig.~\ref{Fig:nsrp}, that the 68\% CL upper limit is $m_\phi < 0.33 \sqrt{\lambda} M_P = 1.4 \times 10^{12}$~GeV (for $y \ge 10^{-15}$ 
and a 95\% CL upper limit of 
$m_{\phi}\lesssim 0.38 \sqrt{\lambda} M_P = 1.6 \times 10^{12}$~GeV. For larger masses it becomes impossible to simultaneously satisfy the {\em Planck} constraints to 2$\sigma$ and the BBN bound $T_{\rm RH}\gtrsim {\rm MeV}$.

In addition to an upper bound to $m_\phi$, we expect that radiative corrections to the potential will provide finite mass which unless fine-tuned away, will determine a lower bound on the inflaton mass.
We expect that through the coupling of the inflaton to either fermions or scalars, would lead to a mass term proportional to $y m_f$ or $\mu$ (see Eqs.~(\ref{eq:ydef}) and (\ref{mu}) for couplings to fermions and scalars respectively. 
While the former is probably no larger than the weak scale, the coupling to scalars could generate a significant contribution to $m_\phi$.  Furthermore, in a supersymmetric theory we would also expect contributions to the scalar mass of order the supersymmetry breaking scale. However, as noted, any lower limit to the inflaton mass would be subject to the degree of fine-tuning by canceling a bare mass term with any 1-loop corrections. Therefore unlike the upper limit discussed above, we do not apply a firm lower limit its mass, but recognize that it should not be surprising to generate weak scale masses, even in theories with the potential given in Eq.~(\ref{Vatt}) for $k\ge4$.

\section{Consequences of the inflaton coupling to matter}
\label{Sec:coupling}

Reheating to create a thermal bath of Standard Model particles requires some coupling of the inflaton to the Standard Model. 
The relation between this coupling and the reheating temperature is dependent not only on the shape of the inflaton potential about its minimum, 
but also on whether the reheating is produced by inflaton decay (in to either fermions, scalars or vectors) or scattering.  As in \cite{GKMO1,GKMO2} we 
will study the three possible cases: fermion decay, scalar decay and scalar scattering, adding the vectorial final states (decay and scattering) analyzed in \cite{gkkmov}.

\subsection{Inflaton decay to fermions}

Given a Yukawa-like coupling of the inflaton to 
fermions,
\beq\label{eq:ydef}
{\cal L}_{\phi ff}=y\phi \bar f f\,,
\eeq
the inflaton decay rate is
\beq
\Gamma_\phi=\frac{y_{\rm eff}^2}{8 \pi} {m_\phi} \,.
\eeq
Here, the effective Yukawa coupling $y_{\rm eff}(k) \neq y$ is defined by averaging over an oscillation. 
In general for $k\ne 2$, the effective coupling must be calculated numerically \cite{GKMO2,Shtanov:1994ce,Ichikawa:2008ne}.

The general expressions for the reheating temperature, defined by $\rho_\phi (\arh) = \rhor (\arh)$
and $\alpha \trh^4 = \rhor(\arh)$, are given in the Appendix.
$\trh$ depends strongly on the spin of the final state decay products, and for decays to fermions, Eq.~(\ref{trhpos}) gives with $l = 1/2-1/k$ and $k<7$
\beq
T_{\rm RH}=\left(\frac{1}{\alpha} \right)^{\frac{1}{4}}
\left[\frac{k \sqrt{3k(k-1)}}{7-k} \lambda^{\frac{1}{k}}
 \frac{y_{\rm eff}^2}{8 \pi}\right]^{\frac{k}{4}}M_P \, ,
\label{trf}
\eeq
or
\beq
\trh \; = \; 
\begin{cases}
\label{trh2}
\left(\frac{\lambda}{\alpha}\right)^\frac14 \frac{y_{\rm eff}^2}{\pi} M_P \simeq 4.2\times 10^{14} y_{\rm eff}^2 ~{\rm GeV}  \qquad k=4 \,,\\
\left( \frac{3}{\alpha} \right)^\frac14 \left(\frac{y_{\rm eff}^2 m_\phi M_P}{20 \pi} \right)^\frac12 \\
 \qquad \simeq 3.3\times 10^{12} y_{\rm eff} \sqrt{\frac{m_\phi}{10^9 {\rm GeV}}} ~{\rm GeV}
 \qquad k=2 \, .
\end{cases}
\eeq
Notable in Eq.~(\ref{trh2}) is that  $\trh$ exhibits a different dependence on the coupling and mass of the inflaton.
In particular,  $\Trh\propto y_{\rm eff}^2$
in the case $\am>\arh$, $\Trh\propto y _{\rm eff} \sqrt{m_\phi}$ if $\am<\arh$.
We will see that for sufficiently low coupling, the quadratic term can dominate the reheating process leading to a higher reheating temperature.

When the limit in Eq.~(\ref{Eq:limittrh}) is satisfied, reheating is sufficiently late to be determined by the 
quadratic term ($k=2$ in Eq.~(\ref{trh2})) and that can be translated into a limit on the coupling $y_{\rm eff}$,
\beq
y_{\rm eff} \lesssim y_{\rm eff}^{\rm m}= 0.02 \sqrt{\frac{m_\phi}{10^{9}~\rm{GeV}}}\,.
\label{Eq:limity}
\eeq

\begin{figure}[!ht]
\centering
\includegraphics[width=\columnwidth]{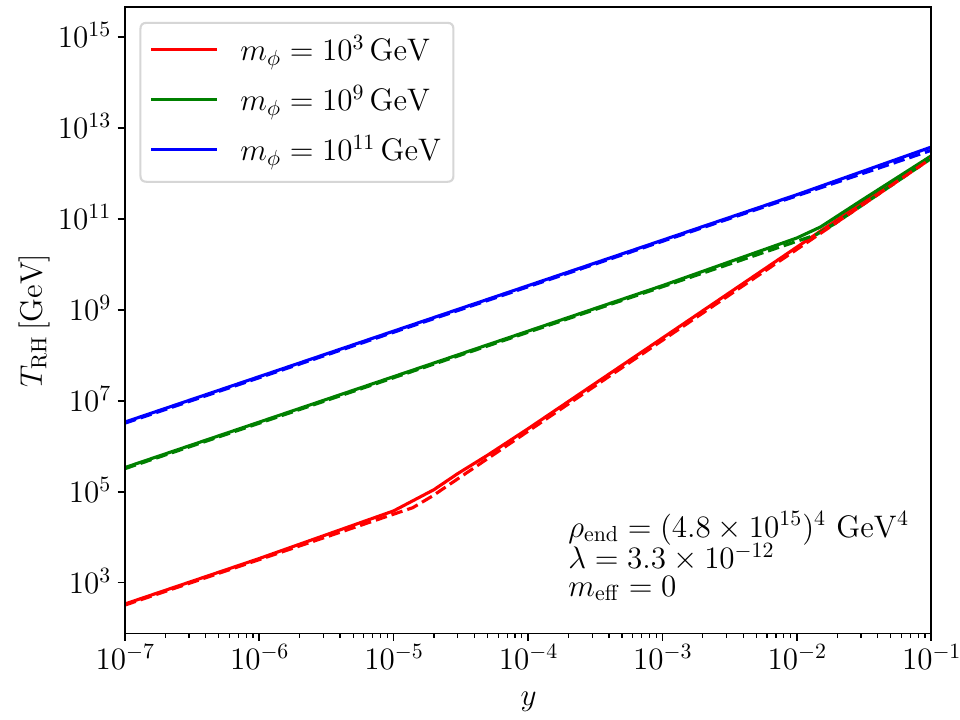}
\caption{\em \small Reheating temperature as a function of the Yukawa coupling $y$ when a bare mass term is added to a quartic potential ($k=4)$. Solid lines are obtained by solving numerically the Boltzmann equations for energy densities, while dashed lines are given by the analytical approximations in Eqs. (\ref{trh2}) Here we neglect the effective mass of the final state  fermion, $\mathcal{R} = 0$ and $y_{\rm eff} = y$. 
 }
\label{Fig:ploty_num}
\end{figure}

We show in Fig.~\ref{Fig:ploty_num} the value of the reheating temperature as function of $y_{\rm eff}$ for different values
of the inflaton bare mass $m_\phi=10^3, ~10^9$ and 
$10^{11}$ GeV, neglecting the effects of an effective final state mass (see below)
and thus $y_{\rm eff} = y$.
To obtain the figure, we solved 
numerically the complete set of Friedmann equations
for $\rho_R$ and $\rho_\phi$, taking the full potential
$V(\phi)=\frac{1}{2}m_\phi^2 \phi^2 + \lambda \phi^4$.
We also show for comparison with dashed lines, the analytical value of $\trh$ obtained
in Eqs.~(\ref{trh2}).
We clearly see the change of behavior $\Trh=f(y_{\rm eff})$ below the limiting value in Eq.~(\ref{Eq:limity})
where the bare mass term controls the final reheating temperature. For $y_{\rm eff} \lesssim y_{\rm eff}^{\rm m}$, $\trh \propto y_{\rm eff}$, whereas for larger 
values of $y_{\rm eff}$, when the reheating is 
dominated by the quartic part of the potential, 
the reheating temperature $\propto y_{\rm eff}^2$ and is independent of $m_\phi$. 

A background field value for $\phi$, however, 
induces an effective mass for the fermion, $f$, $m_{\rm eff} = y \phi$, and the rates for producing the fermions are 
suppressed by ${\cal R}^{-1/2}$ where ${\cal R} \propto m_{\rm eff}^2/m_\phi^2 \propto y^2 (\phi_0/M_P)^{4-k}/\lambda$ \cite{GKMO2}. 
The mass of the inflaton is defined by
\beq
\label{eq:mephiff}
m_{\phi}^2(t) \;\equiv\; V''(\phi_0(t)) \,.
\eeq
When ${\cal R} \gg 1$, there is a significant suppression in the decay rate and $y_{\rm eff} \ll y$. 
Note that in the case of a quartic potential, 
$m_\phi\propto \phi$. As $m_{\rm eff}\propto \phi$
also, ${\cal R}$ is constant, ${\cal R} \simeq 1.4 y^2/\lambda \simeq 4.2 \times 10^{11} y^2$. In other words, the effect of ${\cal R}$ results in a suppression of the reheating efficiency by a constant factor 
${\cal R}^{-\frac{1}{2}}\simeq 1.5\times 10^{-6}/y$  throughout the reheating process. This suppression begins to be efficient 
(${\cal R}\gtrsim 1$) for $y \gtrsim 1.5\times 10^{-6}$
\cite{GKMO2}.
On the other hand, for a quadratic potential, ${\cal R} =4 (\phi_0/m_\phi)^2 y^2$
{\it decreases} with time, redshifting as $a^{-\frac{3}{2}}$. Which means that if there is no suppression
during the quartic dominated era ($a<\am$), there is no suppression in the quadratic era ($a>\am$). 

The kinematic suppression in the effective coupling $y_{\rm eff}$ for $\mathcal{R}\gg 1$ can be parametrized as~\cite{GKMO2}
\beq
 y_{\rm eff}^2 \simeq c_k {\mathcal{R}}^{-1/2} (\omega/m_\phi) y^2
\eeq
where $c_k$ is a $k$-dependent constant\footnote{There is an additional dependence of $y_{\rm eff}$ on the sum of the Fourier modes associated with the inflaton oscillations in the potential $V(\phi)\sim \phi^k$, for each value of $k$. However, this additional dependence is $\mathcal{O}(1)$, as shown in \cite{GKMO2}.} and $\omega$ is the oscillation frequency. For $k=4$, $c_4 \simeq 0.5$ and $\omega \simeq 0.49 m_\phi$. This leads to 
\beq
y_{\rm eff}\simeq \frac12 \times \frac{y}{{\cal R}^\frac{1}{4}}
\simeq 6\times 10^{-4} \sqrt{y} \qquad (k=4) \, .
\label{Eq:yeff}
\eeq
We note that only when ${\cal R} \sim 0.1$, do we recover $y_{\rm eff} = y$.  Note also that unless $y_{\rm eff}$ is relatively small, $y_{\rm eff} \lesssim 2\times 10^{-3}$, the Lagrangian coupling, $y$, is non-perturbative \cite{GKMO2,GGMOPY}, where this perturbativity limit on $y_{\rm eff}$ assumes $y \lesssim \sqrt{4 \pi}$.

For $k=2$, $c_2 \simeq 0.38$ and $\omega = m_\phi$. At the end of reheating, 
$\rhorh  = \frac12 m_\phi^2 \phi_0^2(\arh) = \alpha \trh^4$, so that
\beq
\phi_0(\arh) = \sqrt{2\alpha}\frac{\trh^2}{m_\phi} \, .
\label{phi0tr}
\eeq
Then, for $\mathcal{R}\gg 1$, we can write
\beq
y_{\rm eff} \simeq 0.15 (m_\phi/\trh) \sqrt{y} \qquad (k=2) \, ,
\eeq
and using Eq.~(\ref{trh2}) for $\trh$ in terms of $y_{\rm eff}$ we have 
\beq
y_{\rm eff} = 6.7 \times 10^{-3} \left(\frac{m_\phi}{10^9~\rm GeV}\right)^\frac14 y^\frac14 \qquad (k=2) \, .
\label{yeffy2}
\eeq
In this case, non-
perturbativity sets in unless $y_{\rm eff} \lesssim 1.5 (m_\phi/\phi_0)^\frac12$, assuming that $y_{\rm eff} < y$. 
Note that for $k > 4$ the limit becomes more severe as $\mathcal{R}$ is larger and increases in time.

\begin{figure}[!ht]
\centering
\includegraphics[width=\columnwidth]{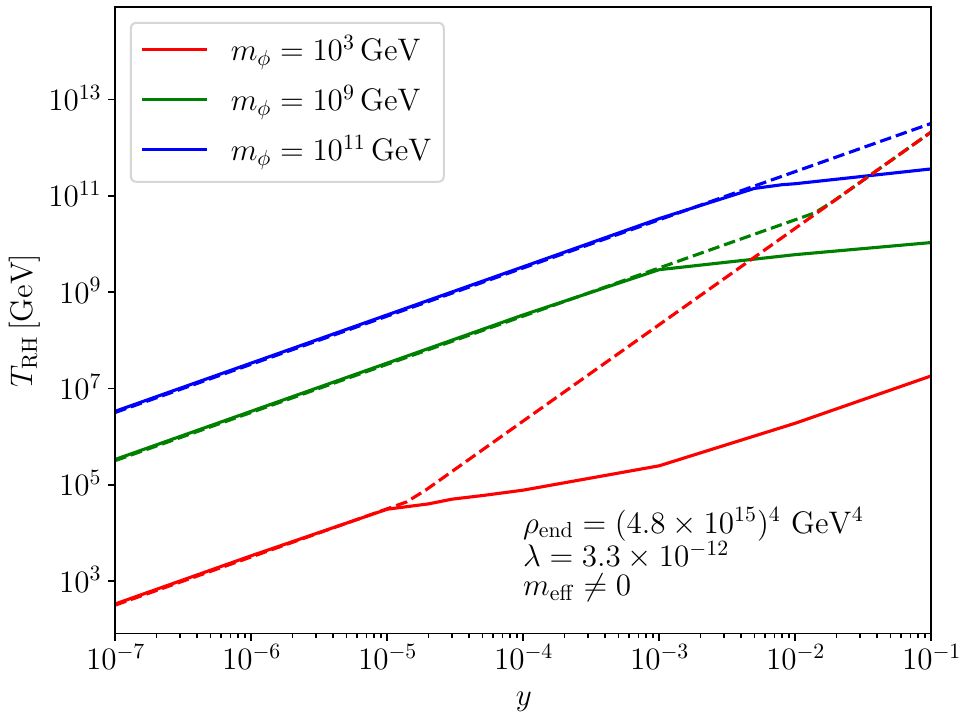}
\caption{\em \small As in Fig.~\ref{Fig:ploty_num}, the reheating temperature as function of the yukawa coupling $y$ 
for different values of the inflaton bare mass $m_\phi=10^3$ GeV (red-dotted), $10^9$ GeV (green-dashed) and $10^{11}$ GeV (full-blue). Here we consider the effective mass of produced fermion, $\mathcal{R} =(2y\phi_0/\omega)^2$.  
}
\label{Fig:ploty_num_R}
\end{figure}

Because of the suppression in the decay rate,
the relation between $\Trh$ and the decay coupling $y$ shown in Fig.~\ref{Fig:ploty_num} needs to be reassessed. Indeed, when $y \gtrsim 1.5 \times 10^{-6} $, ${\cal R} \gtrsim 1$ and the suppression effect should be taken into account. The relation between $\trh$ and $y$
when the effects of kinematic suppression are included is shown in Fig.~\ref{Fig:ploty_num_R}. 
At very low values of $y$, ${\cal R} \ll 1$ and the suppression effects can be ignored. In this case, the relation between $\trh$ and $y$ is unaffected.   However, when $y_{\rm eff} \le y$ the relation is altered. 
From Eq.~(\ref{yeffy2}), this occurs when $y > 1.3 \times 10^{-6} (m_\phi/{\rm GeV})^\frac13$, or when $y > 1.3 \times 10^{-5} (1.3 \times 10^{-3}) (6\times 10^{-3})$ when $m_\phi = 10^3~(10^9)~(10^{11})$~GeV. 
These values are seen in Fig.~\ref{Fig:ploty_num_R} when the solid curves begin to deviate from the dashed curves. The dashed curves show the relation in Fig.~\ref{Fig:ploty_num} when suppression effects are ignored. 
The expression for $y_{\rm eff}$ in Eq.~(\ref{yeffy2}) can be inserted in Eq.~(\ref{trh2}) to obtain the relation between $\trh$ and $y$ for when suppression effects are included and reheating is governed by the quadratic term,
\beq
\trh = 2.2 \times 10^{10}~{\rm GeV} \left(\frac{m_\phi}{10^9 {\rm GeV}} \right)^\frac34 y^\frac14 \qquad (k=2) \, .
\eeq

We saw previously in Eq.~(\ref{Eq:limittrh}) that the reheating temperature is determined by the quartic term only if $\trh \gtrsim 250 m_\phi$. When the kinematic suppression effects are ignored ($y = y_{\rm eff}$),
this occurs when $y$ does not satisfy Eq.~(\ref{Eq:limity}). In this case, we can 
use Eq.~(\ref{Eq:yeff}) to determine the relation between $\trh$ and $y$,
\beq
\trh = 1.5 \times 10^8 y ~{\rm GeV} \qquad (k=4) \, ,
\eeq
and thus we expect that reheating is determined by the quartic term when $y > 1.7 \times 10^{-6} m_\phi/{\rm GeV}$. This occurs at $y = 1.7 \times 10^{-3}$
for $m_\phi = 10^3$~GeV as can be seen in Fig.~\ref{Fig:ploty_num_R}. For the larger masses shown, we see that the transition would only occur in the non-perturbative regime (with $y \gg 1$)
and so for the two higher masses, the reheating temperature is always determined by the quadratic mass term.

\subsection{Decay to scalars}

Another possibility is that reheating occurs predominantly through inflaton decay to scalars, through the coupling 
\beq
\mathcal{L}_{\phi b^2} = \mu \phi b^2
\label{mu}
\eeq
where $b$ is a real scalar field. 
As was the case for the fermion decay, there is also an 
effect from the effective mass of the scalar field, 
and we parameterize it by considering an effective coupling $\mu_{\rm eff}$. We note that $\mu_{\rm eff}$ is now a dimensionful parameter and is {\it enhanced}
(and not reduced) by ${\cal R}^{1/2}$ 
\cite{GKMO2}. The associated decay rate is given by 
\beq
\Gamma_{\phi b^2}=\frac{\mu_{\rm eff}^2}{8\pi m_\phi} \, .
\eeq
For $k=2$ this effective coupling reduces to the Lagrangian coupling $\mu$ but is different for $k>2$. It is important
to note that in this case, as $m_\phi$ decreases with time, 
the decay rate {\it increases} with time.

For decays to scalars, $l = 1/k - 1/2$, and using the appropriate expression found in the Appendix for $\gamma_\phi$, we have
\beq
T_{\rm RH}=\left(\frac{1}{\alpha} \right)^{\frac{1}{4}}
\left[\frac{2k \sqrt{3}}{(4k+2)\sqrt{k(k-1)}} \lambda^{-\frac{1}{k}}
 \frac{\mu_{\rm eff}^2}{8 \pi M_P^2}\right]^{\frac{k}{4(k-1)}}M_P \, ,
\label{trb}
\eeq
or
\beq
\trh \; = \; 
\begin{cases}
\label{trh2mu}
\left(\frac{1}{\alpha}\right)^\frac14 \left(\frac{\mu_{\rm eff}^2}{36\pi M_P^2}\right)^\frac13 \lambda^{-\frac{1}{12}} M_P \\
\qquad \simeq 1.8\times 10^{18} \left(\frac{\mu_{\rm eff}}{M_P}\right)^{\frac23} {\rm GeV}  \qquad k=4 \,,\\
\left( \frac{3}{\alpha} \right)^\frac14 \left(\frac{M_P}{20 \pi m_\phi} \right)^\frac12 \mu_{\rm eff} \\
 \qquad \simeq 3.3\times 10^{3} \mu_{\rm eff} \sqrt{\frac{10^9 {\rm GeV}}{m_\phi}} 
 \qquad k=2 \, ,
\end{cases}
\eeq

\begin{figure}[!ht]
\centering
\includegraphics[width=\columnwidth]{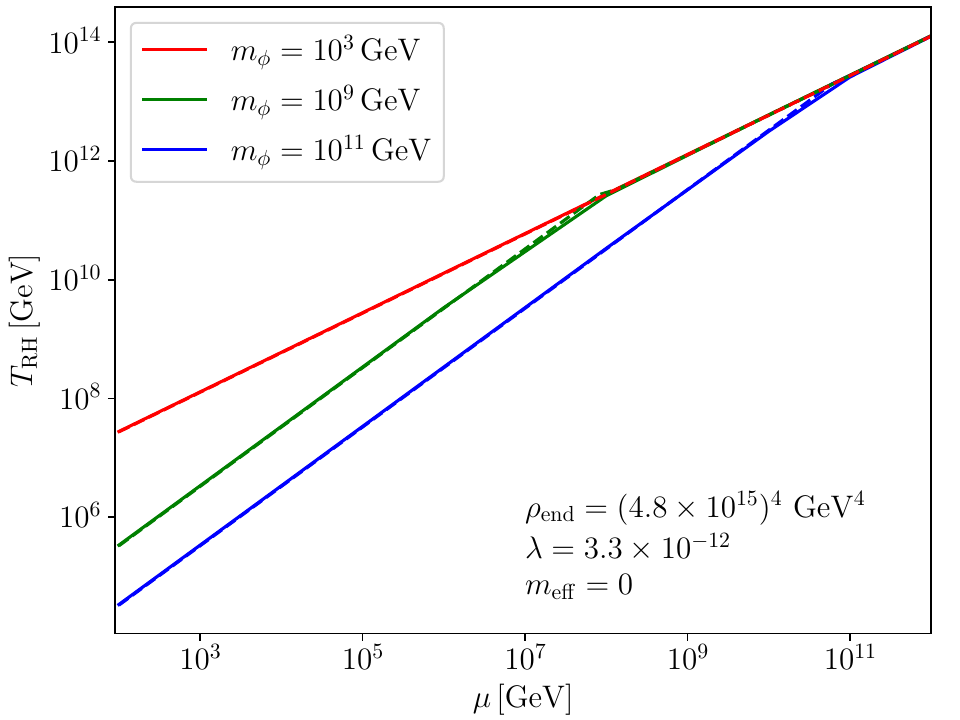}
\caption{\em \small Reheating temperature as function of the bosonic coupling $\mu$,
for different values of the inflaton bare mass $m_\phi=10^3$ GeV (red), $10^9$ GeV (green) and $10^{11}$ GeV (blue). Solid lines are obtained by solving numerically the Boltzmann equations for energy densities, while dashed lines are given by the analytical approximations in Eqs. (\ref{trh2mu}). Here we neglect the effective mass of produced bosons, $\mathcal{R} =  0$.}
\label{Fig:plot_num_mu}
\end{figure}

We show in Fig.~\ref{Fig:plot_num_mu} the evolution of
$\trh$ as function of $\mu$ for the same set of masses
$m_\phi=10^3$, $10^9$ and $10^{11}$ GeV, in the simplified case with $m_{\rm eff}=0$. We clearly recognize the dependence 
$\trh \propto \mu$ for the smaller values of $\mu$ and
$\trh \propto \mu^{2/3}$ for the larger values, when
reheating is dominated by the quartic part of the potential.
The value of $\mu$ for which  
reheating is dominated by the quadratic term obtained from Eq.~(\ref{trh2mu}) with $k=4$ is
\beq
\mu \lesssim 1.3\times 10^8
\left(\frac{m_\phi}{10^9~\rm{GeV}}\right)^\frac{3}{2}~\rm{GeV}\,,
\label{Eq:limitmur0}
\eeq
which is effectively what is observed in Fig.~\ref{Fig:plot_num_mu}. 
From Eq.~(\ref{Eq:limitmur0}), we see that the reheating temperature for $m_\phi = 10^3$~GeV (red curve) is always due to the quartic term, as the transition from quadratic to quartic occurs at a low value of $\mu$ beyond the range shown. 
For the larger values of $m_\phi$, Eq.~(\ref{Eq:limitmur0}) indicates when  when the slopes of $\trh$ vs. $\mu$ begins to change.

\begin{figure}[!ht]
\centering
\includegraphics[width=\columnwidth]{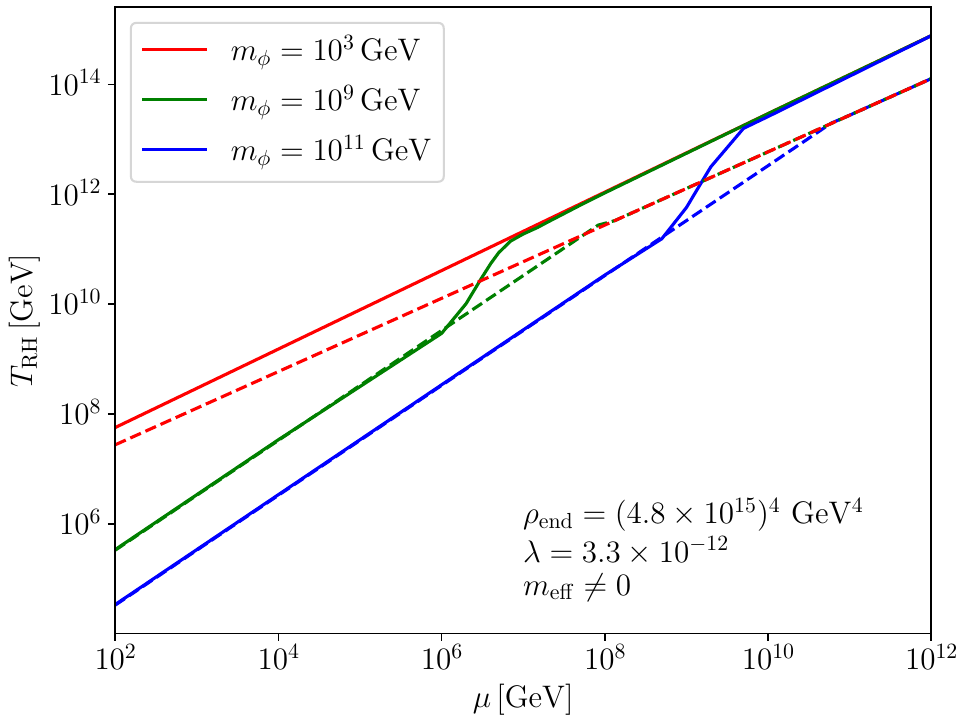}
\caption{\em \small Reheating temperature as function of the bosonic coupling $\mu$,
for different values of the inflaton bare mass $m_\phi=10^3$ GeV (red), $10^9$ GeV (green) and $10^{11}$ GeV (blue). Solid lines are obtained by solving numerically the Boltzmann equations for energy densities, while dashed lines are given by the analytical approximations in Eqs. (\ref{trh2mu}). Here we consider the effective mass of produced bosons, $\mathcal{R} = 8\mu\phi_0/\omega^2$.}
\label{Fig:plot_num_mu_Rneq0}
\end{figure}

In order to account for the effective
mass $m_{\rm eff}^2 = 2 \mu \phi_0$, we need to include an {\it enhancement} of the production rate 
$\propto {\cal R}^{\frac{1}{2}}$, with ${\cal R}=8 \mu \phi_0/m_\phi^2$ for $k=2$ and ${\cal R} \simeq 2.8 \mu/(\lambda \phi_0)$ for $k=4$. 
The effective dimensionful coupling\footnote{Again, an additional $\mathcal{O}(1)$ dependence of $\mu_{\rm eff}$ on the sum of the Fourier modes associated with the inflaton oscillations for each value of $k$ is neglected here \cite{GKMO2}. Note also that the values of $c^\prime_k$ were omitted in \cite{GKMO2}.} when $\mathcal{R}\gg1$ is \cite{GKMO2} 
\beq
\mu_{\rm eff}^2 \simeq \frac{c_k'}{4} (k+2)(k-1) \frac{\omega}{m_\phi}{\cal R}^\frac12 \mu^2 \, ,
\label{mueff}
\eeq
with $c_k'\simeq\{0.38,0.37,0.36\}$ for $k=\{2,4,6\}$,
so that $\mu_{\rm eff} \simeq 0.62 (8 \phi_0/m_\phi^2)^\frac14 \mu^\frac54$ for $k=2$.
Then using Eq.~(\ref{phi0tr}) for $\phi_0$ and Eq.~(\ref{trh2mu}) for $k=2$ to replace $\trh$, we have
\beq
\mu_{\rm eff} \simeq 3.3\times 10^{-10} ~{\rm GeV} \left(\frac{10^9~{\rm GeV}}{m_\phi} \right)^2 \left(\frac{\mu}{\rm GeV} \right)^\frac52 \, .
\label{mueff2}
\eeq
Then, the effects of the kinematic enhancement will occur when
\beq
\mu \gtrsim 2.1 \times 10^6 \left(\frac{m_\phi}{10^9~\rm{GeV}}\right)^\frac{4}{3}\,.
\eeq
This can be see seen in Fig.~\ref{Fig:plot_num_mu_Rneq0} for $m_\phi = 10^{9}~(10^{11}$~GeV as the point when the solid curves break away from the dashed curves at $\mu \simeq 2.1 \times 10^6~(9.8 \times 10^8)$~GeV respectively. At lower values of $\mu$,
the effects of the kinematic suppression can be ignored. For $m_\phi = 10^3$~GeV, this occurs at a value of $\mu$ below the range shown. 

In the region when $\mu_{\rm eff} > \mu$ and quadratic reheating dominates, we can insert Eq.~(\ref{mueff2}) into Eq.~(\ref{trh2mu}) to obtain
\beq
\trh = 1.1 \times 10^{-6}~{\rm GeV}~\left(\frac{10^9~{\rm GeV}}{m_\phi} \right)^\frac52 \mu^\frac52 \qquad (k=2) \, .
\eeq

At higher values of $\mu$ the transition to quartic reheating occurs and using Eq.~(\ref{mueff}) with the expression for ${\cal R}$ for $k=4$, 
we find that 
\beq
\mu_{\rm eff} \simeq 2.5~{\rm GeV} \left(\frac{\mu}{\rm GeV} \right)^\frac{15}{14}
\eeq
which when inserted in Eq.~(\ref{trh2mu})
gives
\beq
\trh \simeq 2.5\times 10^{19}~{\rm GeV} \left(\frac{\mu}{M_P}\right)^\frac57 \qquad (k=4) \, .
\eeq

\subsection{Decay to Vectors}

Recently, we have considered the possibility of inflaton decays to vectors
\cite{gkkmov} motivated by inflationary 
models in the context of no-scale supergravity \cite{building} (which easily lend construction of the $T$-models considered here \cite{GKMO1}).
Often in such models, the inflaton couplings to matter fermions and scalars are highly suppressed \cite{ekoty,egno4,enov4} and reheating is only possible if the gauge kinetic functions contain inflaton couplings.
The inflaton to vector couplings can be parameterized by
\begin{equation}
    \mathcal{L}\supset -\frac{g}{4M_P}\phi F_{\mu\nu}F^{\mu\nu}-\frac{\Tilde{g}}{4M_P}\phi F_{\mu\nu}\Tilde{F}^{\mu\nu} \, ,
    \label{gaugekincouplings}
\end{equation}

From these Lagrangian couplings, we can derive the inflaton decay rate
\beq
 \Gamma_{\phi\rightarrow A_\mu A_\mu} \; = \;
 \frac{\alpha_{\rm eff}^2 m_\phi^3}{M_P^2}\,,
 \label{Eq:gammadecay}
\eeq
where $\alpha^2_{\rm eff} = (g_{\rm eff}^2 + \tilde g_{\rm eff}^2)/(64\pi)$. Note the dependence 
of the width on $m_\phi^3$, which is very different from the decay into fermions ($\propto m_\phi$) and to scalars ($\propto 1/m_\phi$).
$\Gamma_{\phi \rightarrow A_\mu A_\mu}$ decreases much more rapidly than 
$\Gamma_{\phi \rightarrow ff}$, rendering the reheating much less efficient,
even impossible as long as the reheating is dominated by the quartic term.

Indeed, for decay to vectors, $l = 3/2 - 3/k$, and using the appropriate expression for $\gamma_\phi$, we have from Eq.~(\ref{trhpos})
\beq
T_{\rm RH}=\left(\frac{1}{\alpha} \right)^{\frac{1}{4}}
\left[\frac{ \sqrt{3} k^\frac52 (k-1)^\frac32 \lambda^{\frac3k}}{13-4k} 
\alpha_{\rm eff}^2\right]^{\frac{k}{4(3-k)}}M_P \, .
\label{trA}
\eeq
This expression is valid so long as $k+8-6kl > 0$, which is the case for $k=2$, but not for $k\ge4$. For $k+8-6kl < 0$, the reheating temperature is given by Eq.~(\ref{trhneg}) for $k>4$. For $k=4$, the radiation density in Eq.~(\ref{Eq:rhoRappter}) scales as $a^{-4}$ as does the inflaton energy density in Eq.~(\ref{Eq:rhophi}) and we never achieve the condition that $\rho_\phi(\arh) = \rho_R(\arh)$ and reheating never occurs. 
Thus we have
\beq
\trh \; = \; 
\begin{cases}
\label{trh2al}
{\rm no~reheating} \qquad \qquad k=4 \,,\\
\left( \frac{3}{\alpha} \right)^\frac14 \left( \frac{2 m_\phi^3}{5 M_P^3} \right)^\frac12 \alpha_{\rm eff} M_P \\
 \qquad \simeq 7.0\times 10^{3} \alpha_{\rm eff} \left(\frac{m_\phi}{10^9 {\rm GeV}}\right)^\frac32 ~{\rm GeV}
 \qquad k=2 \, ,
\end{cases}
\eeq

Thus for a $k=4$ inflationary potential, reheating via the decays to vector bosons does not occur in the absence of a bare mass term.  The bare mass term
is then {\it necessary} to ensure a successful reheating. 
However, the bare mass term should ensure $\trh \gtrsim 2$ MeV, 
which means 
\beq
m_\phi \gtrsim 40 \alpha_{\rm eff}^{-\frac23}~{\rm TeV} \, .
\eeq
This value is the minimal bare mass necessary to have reheating 
through decay to vectors for $k=4$.

Finally we note that there are no kinematic enhancement/suppression
effects in this case.  Since the inflaton is coupled to $F^2$ (as opposed to $A^2$), no mass term is generated. Then $g_{\rm eff} = g$ (and $\tilde g_{\rm eff}= \tilde g$) for $k=2$, and for $k=4$ only differs by a Fourrier coefficient in an expansion of $V(\phi)$ \cite{gkkmov}.
 
\subsection{Scattering to scalars}
We can also consider the case where the inflaton transfers its energy through the coupling 
\beq
\mathcal{L}_{\phi^2 b^2} = \sigma \phi^2 b^2
\eeq
where $b$ is a real scalar field. The associated decay rate is given by \cite{GKMO2}
\beq
\Gamma_{\phi^2 b^2} = \frac{\sigma_{\rm eff}^2}{8\pi}\frac{\rho_\phi}{m_\phi^3}
\eeq
where we have introduced the effective coupling $\sigma_{\rm eff}$ obtained, as for $y_{\rm eff}$ and $\mu_{\rm eff}$, after averaging over oscillations of the background inflaton condensate \cite{GKMO2}. This effective coupling is equal to the Lagrangian coupling $\sigma$ for $k=2$ but is different for $k>2$ and as in the case of decays to fermions there is a kinematic suppression. 

For scattering to scalars, $l = 3/k - 1/2$, and using the appropriate expression found in the Appendix for $\gamma_\phi$, we have
from Eq.~(\ref{trhpos}) valid when $k\ge4$,
\beq
T_{\rm RH}=\left(\frac{1}{\alpha} \right)^{\frac{1}{4}}
\left[\frac{ \sqrt{3}}{(2k-5)\sqrt{k}(k-1)^\frac32} \lambda^{-\frac{3}{k}}
 \frac{\sigma_{\rm eff}^2}{8 \pi }\right]^{\frac{k}{4(k-3)}}M_P \, .
\label{trss}
\eeq
For $k=2$, $8+k-6kl < 0$ and $\rho_R$
redshifts as $a^{-4}$ which is faster than $\rho_\phi \propto a^{-3}$. Thus, in this case, reheating is not possible if the quadratic term becomes dominant before reheating is complete. 
The reheating temperature can then be written as,
\beq
\trh \; = \; 
\begin{cases}
\label{trh2s}
\left(\frac{1}{\alpha}\right)^\frac14 \left(\frac{\sigma_{\rm eff}^2}{144\pi}\right)\lambda^{-\frac{3}{4}} M_P \\
\qquad \simeq 8.9\times 10^{23} \sigma_{\rm eff}^2 {\rm GeV}  \qquad k=4 \,,\\
{\rm no~reheating} \qquad
 \qquad k=2 \, .
\end{cases}
\eeq

As one can see, the possibility of reheating through scattering to scalars is opposite the case of decays to vectors. 
Reheating is {\it not possible} when the quadratic part of the potential dominates the reheating process.
Naively, when we neglect the kinematic suppression effects in $\cal R$, reheating is therefore only possible if the limit in Eq.~(\ref{Eq:limittrh}) is violated, namely
\beq
\sigma_{\rm eff} \gtrsim 5.3\times 10^{-7} \sqrt{\frac{m_\phi}{10^9~\rm GeV}} \, .
\label{Eq:sigmaeff}
\eeq
For smaller couplings, the quadratic term will dominate before reheating is complete, and as a result never completes. 
We note in the expression for $\trh$ in Eq.~(\ref{trh2s}), the maximum value for $\sigma_{\rm eff}$ that can be used is determined from  $\arh > \aend$, which gives 
\beq
\sigma_{\rm eff}^2 < 2.2 \times 10^{-9} \, .
\label{siglim}
\eeq
Furthermore for self couplings this large, we expect that non-perturbative effects become non-negligible \cite{GKMOV}.
For larger values, we have a maximum reheating temperature of $2 \times 10^{15}$~GeV, which is basically determined from $\rhoe$. 

As previously noted, for inflaton scattering to scalars, there is a kinematic suppression when ${\cal R} > 1$. In this case, for $k=4$, ${\cal R} \simeq 2.8 \sigma/\lambda$ is a constant
and \footnote{We neglect the dependence of $\sigma_{\rm eff}$ on the sum of the Fourier modes associated with the inflaton oscillations for each value of $k$ \cite{GKMO2}.}
\begin{eqnarray}
\sigma_{\rm eff}^2 &\simeq & \frac{c^{\prime\prime}_k}{8} k (k+2)(k-1)^2 {\mathcal{R}}^{-1/2} (\omega/m_\phi) \sigma^2 \nonumber \\
& \simeq & 16 {\mathcal{R}}^{-1/2} \sigma^2 \simeq 9.6 \sqrt{\lambda} \sigma^\frac32 \, ,
\label{seff2}
\end{eqnarray}
using $c^{\prime\prime}_4 = 1.22$.
Then the reheating temperature in terms of $\sigma$ becomes
\beq
\trh = 1.6 \times 10^{19}~{\rm GeV} \sigma^\frac32 \qquad (k=4) \, .
\label{trhs2}
\eeq
In Fig.~\ref{Fig:plot_num_sigma}, we compare the reheating temperature as a function of $\sigma$ when kinematic effects are ignored
to the case where they are included. From Eq.~(\ref{seff2}), these effects become important when $\sigma > 3.1 \times 10^{-10}$. The dashed lines correspond to the solution when kinematic effects are ignored. The abrupt increase in $\trh$ occurs when Eq.~(\ref{Eq:sigmaeff}) is satisfied  (and $\sigma_{\rm eff} = 4 \sigma$). In contrast, the solid lines include the kinematic suppression and reheating is possible when Eq.~(\ref{Eq:sigmaeff}) is used with Eq.~(\ref{seff2}) or when
\beq
\sigma \gtrsim 6.4 \times 10^{-6} \left( \frac{m_\phi}{10^9~{\rm GeV}} \right)^\frac23 \, .
\eeq
This limit accounts for the abrupt rise in $\trh$ for the solid lines in Fig.~\ref{Fig:plot_num_sigma}. At higher coupling, the reheating temperature follows Eq.~(\ref{trhs2}) and scales as $\sigma^\frac32$ as opposed to $\sigma^2$ when the suppression effects are ignored. In the latter case, we see the curves flatten at large coupling since $\arh$ is approaching $\aend$
and the approximation used in (\ref{trh2s}) breaks down.  These curves end when $\arh = \aend$, indicated by the vertical gray dotted line. The solid curves would end when $\sigma \simeq 0.002$.

\begin{figure}[!ht]
\centering
\includegraphics[width=\columnwidth]{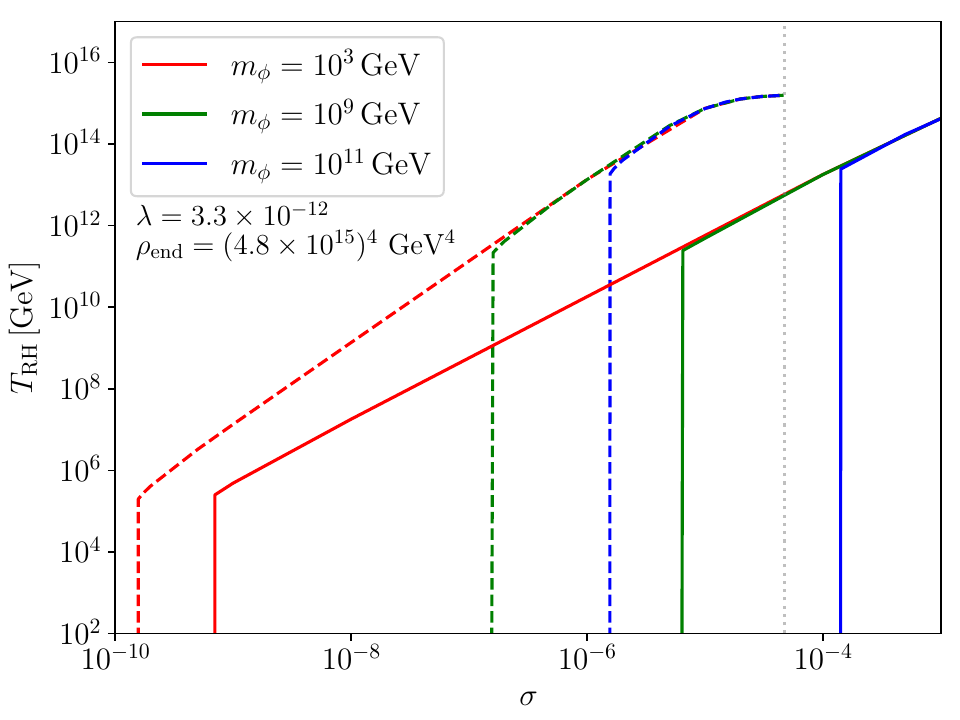}
\caption{\em \small Reheating temperature as function of the scattering coupling $\sigma$,
for different values of the inflaton bare mass $m_\phi=10^3$ GeV (red), $10^9$ GeV (green) and $10^{11}$ GeV (blue). Solid lines are obtained by solving numerically the Boltzmann equations for energy densities including the effect of $\mathcal{R}$, while dashed lines neglect the effect of effective masses. The vertical gray dotted line corresponds to the limit Eq.(\ref{siglim}), when neglecting the effect of $\mathcal{R}$.}
\label{Fig:plot_num_sigma}
\end{figure}

In the absence of a decay term for the inflaton, a bare mass term will eventually lead to a non-zero relic density of inflatons after annihilations freeze out. 
Indeed, even if $\sigma_{\rm eff}$ is sufficiently large and respects the condition (\ref{Eq:sigmaeff}),
the presence of a quadratic term may dominate the energy budget of the Universe.
Thus we can derive a limit on a combination of the inflaton mass, $\trh$ and the coupling $\sigma$.
Saturating the limit leaves us with the inflaton as a cold dark matter candidate!~\footnote{The possibility of inflaton dark matter in a similar context was considered in \cite{Garcia:2021gsy} where the conditions for freeze-out of a thermal inflaton given. See also \cite{Liddle:2006qz,delaMacorra:2012sb,Mukaida:2014kpa,kamion,Lebedev:2021zdh}.}

Indeed, for $\sigma_{\rm eff}$ sufficiently large to ensure reheating with $k=4$,  for $a > \arh$, the evolution of $\rho_\phi$ is determined from the Boltzmann equation including dissipative effects \cite{GKMO2}
\beq
\frac{d}{da}\left( \rho_\phi a^{\frac{6k}{k+2}} \right) =
- \frac{\gamma_\phi}{aH} \frac{2k}{k+2} \frac{\rho_\phi^{l+1}}{M_P^{4l}}
a^{\frac{6k}{k+2}} \, .
\label{drhoda}
\eeq
and for $k=4$, $l = \frac14$ and $\gamma_\phi$ as given in the Appendix, we find that $\rho_\phi$ scales as
\beq
\rho_\phi= 256 \rhorh\left(\frac{\arh}{a}\right)^8\,.
\label{phi8}
\eeq
Here, we used $H= \sqrt{\rho_R/3M_P^2}$.
In the absence of a mass term, since $\Gamma_\phi \propto \gamma_\phi \rho_\phi^\frac14 \propto a^{-2}$
and after reheating, $H \propto \rho_R^\frac12 \propto a^{-2}$, the ratio $\Gamma/H$ remains constant and the scaling in Eq.~(\ref{phi8}) remains true indefinitely and the density of inflatons becomes negligibly small.

However, when $m_\phi \ne 0$, eventually the mass term dominates over the quartic term (at $a=\am$)
 and we can determine $\am$, when 
$\rho_\phi(\am)=\frac12 m_\phi^2 \phi^2(\am)$,
\beq
\frac{\am}{\arh}=\frac{\rhorh^\frac18 \lambda^\frac18 2^\frac18}{\sqrt{m_\phi}}
 \,,
\eeq
where the inflaton density is given by
\beq
\rho_\phi^{\rm m}=\rho_\phi(\am)=\frac{m_\phi^4}{2 \lambda} \, . 
\eeq
as was previously found in Eq.~(\ref{Eq:rhom}).

For $a > \am$, Eq.~(\ref{drhoda}) can be solved, now with $k=2$ and $l=1$. In the limit that $a \gg \am$, the residual inflaton density is given by 
\beq
\rho_\phi(a) \simeq \rho_\phi^{\rm m}\left(\frac{\am}{a}  \right)^3 \, ,
\label{cdm1}
\eeq
so long as $(m_\phi/M_P) \ll (2 \lambda)^\frac14/3^\frac13 \approx .001$, which is always true given the upper limits on $m_\phi$ discussed in Section \ref{masslim}. Thus the presence of a mass term in the case where reheating is determined by a quartic coupling of the inflaton to scalars (which requires $k>2$), leads automatically to cold dark matter candidate.

Given the inflaton density in Eq.~(\ref{cdm1}), 
it is straightforward to compute the relic density today and in effect set a limit on the inflaton bare mass. Today, 
\beq
\rho_\phi = \frac{8 m_\phi^\frac52 \alpha^\frac38 T_0^3}{(2 \lambda)^\frac58 \trh^32} \xi \, ,
\eeq
where $\xi = (43/427)(4/11) \simeq 0.036$
and relative to the critical density we have
\beq
\Omega_\phi h^2 = 1.6 \left(\frac{m_\phi}{1~{\rm GeV}}\right)^\frac52 \left(\frac{10^{10}~{\rm GeV}}{\trh} \right)^\frac32 
\eeq
and thus 
\beq
m_\phi < 0.35 \left( \frac{\trh}{10^{10}~{\rm GeV}} \right)^\frac35~{\rm GeV} \, ,
\label{strlim}
\eeq
using $\Omega_\phi h^2 < 0.12$.  
This is a remarkably strong limit on a bare mass term for the inflaton if it remains stable.

\subsection{Scattering to Vectors}

If the gauge kinetic function is quadratic in the inflaton, then scattering rather decay to vectors occurs. In this case, 
the inflaton to vector couplings can be parameterized by
\begin{equation}
    \mathcal{L}\supset -\frac{\kappa}{4M_P^2}\phi^2 F_{\mu\nu}F^{\mu\nu}-\frac{\Tilde{\kappa}}{4M_P^2}\phi^2 F_{\mu\nu}\Tilde{F}^{\mu\nu} \, ,
    \label{gaugekincouplings2}
\end{equation}

From these Lagrangian couplings, we can derive the inflaton decay rate
\begin{equation}
 \Gamma_{\phi \phi\rightarrow A_\mu A_\mu}=\frac{\beta^2 \rho_\phi }{M_P^4}m_\phi \,,
 \label{Eq:scattering}
\end{equation}
where $\beta^2 = (\kappa_{\rm eff}^2 + \tilde \kappa_{\rm eff}^2)/(4\pi)$.

For scattering to vectors, $l = 3/2 - 1/k$, and using the appropriate expression for $\gamma_\phi$, we have
from Eq.~(\ref{trhneg})
\begin{align}
T_{\rm RH} \;&=\;  \left(\frac{1}{\alpha} \right)^{\frac{1}{4}} \left[\frac{\sqrt{3}k}{4k-7} 
\beta^2 (k(k-1))^\frac12 \lambda^\frac1k \right]^{\frac{3k}{4k-16}} \nonumber \\
& \times
\left(\frac{\rho_{\rm end}}{M_P^4} \right)^{\frac{4k-7}{4k-16}} M_P\, .
\label{skg4}
\end{align}
since $8+k-6kl<0$ for $k\ge2$. 
However, Eq.~(\ref{skg4}) is only valid for $k > 4$. For $k=2 (4)$, $\rho_\phi \propto a^{-3} (a^{-4})$ while $\rho_R \propto a^{-4}$ for all $k$ and reheating is not possible for $k<6$.
For these specific cases, we then have
\beq
\trh \; = \; 
\begin{cases}
\label{trh2AA}
{\rm no~reheating} \qquad \qquad k=4 \,,\\
 {\rm no~reheating} \qquad
 \qquad k=2 \, ,
\end{cases}
\eeq
In this case, the presence of a bare mass will not change
the lack of reheating through the scattering to vectors.

As a conclusion, whereas in the case of decays to fermions or bosons, the 
presence of a quadratic term only acts on the {\it value} of $\trh$,
decreasing the reheating temperature in the former case, increasing it in the 
latter case, the quadratic term when dominant removes the possibility of reheating through 
scattering to scalars but reopens the possibility of reheating through decay to vectors, but does not allow reheating through the scattering to vectors.

\section{Generalized potentials}
\label{sec:gen}

The inflationary potential may be dominated by higher order terms if $k>4$.  In this section, we generalize some of the arguments made above in the event that the inflationary potential is approximated by
\beq
\frac12 m_\phi^2 \phi_0^2 + \lambda \phi_0^k M_P^{4-k}
\eeq
about its minimum.  
In this case, the general expression for the scale factor when the mass term dominates is given by
\beq
\frac{\am}{\aend}=
 \left(\frac{2 \lambda^\frac2k M_P^\frac{2(4-k)}{k} \rhoe^\frac{k-2}{k}}{m_\phi^2}\right )^{\frac{k+2}{6k-12}}
\,,
\label{Eq:amk}
\eeq
with $\rhoe$ given by Eq.~(\ref{rhoe}) and $\lambda$ by Eq.~(\ref{eq:normlambda}).
Then 
\beq
\rho_\phi(a_m) = 2 \left(\frac{m_\phi^2}{2}\right)^\frac{k}{k-2} \lambda^{\frac{-2}{k-2}} M_P^\frac{2(k-4)}{k-2}\, ,
\eeq
which clearly reduces to Eq.~(\ref{Eq:rhom}) for $k=4$.
A parallel derivation leading to Eq.~(\ref{Eq:limittrh})
implies that
\beq
\rho_{\rm RH} \lesssim 2 \left(\frac{m_\phi^2}{2} \right)^\frac{k}{k-2} \lambda^{\frac{-2}{k-2}} M_P^\frac{2(k-4)}{k-2}
\eeq
for the mass term to dominate at reheating. In terms of the reheating temperature, this amounts to
\beq
\trh \lesssim \left(\frac{1}{\alpha} \right)^{\frac{1}{4}} \left( \frac{m_\phi M_P^{\frac{k-4}{k}}}{(2 \lambda)^\frac1k} \right)^\frac{k}{2(k-2)} \, .
\label{Eq:limittrhgen}
\eeq
For comparison with Eq.~(\ref{Eq:limittrh}), we have 
\beq
\trh \; \lesssim \; 
\begin{cases}
5.0 \times 10^5~{\rm GeV} \left(\frac{m_\phi}{\rm GeV}\right)^\frac34 \qquad k=6 \,,\\
6.3 \times 10^6~{\rm GeV} \left(\frac{m_\phi}{\rm GeV}\right)^\frac23 
 \qquad k=8 \, ,
\end{cases}
\eeq
using $\lambda = 5.7\times 10^{-13}$ and $9.5\times 10^{-14}$ 
for $k = 6$ and 8, respectively. 

\begin{figure}[!ht]
\centering
\includegraphics[width=\columnwidth]{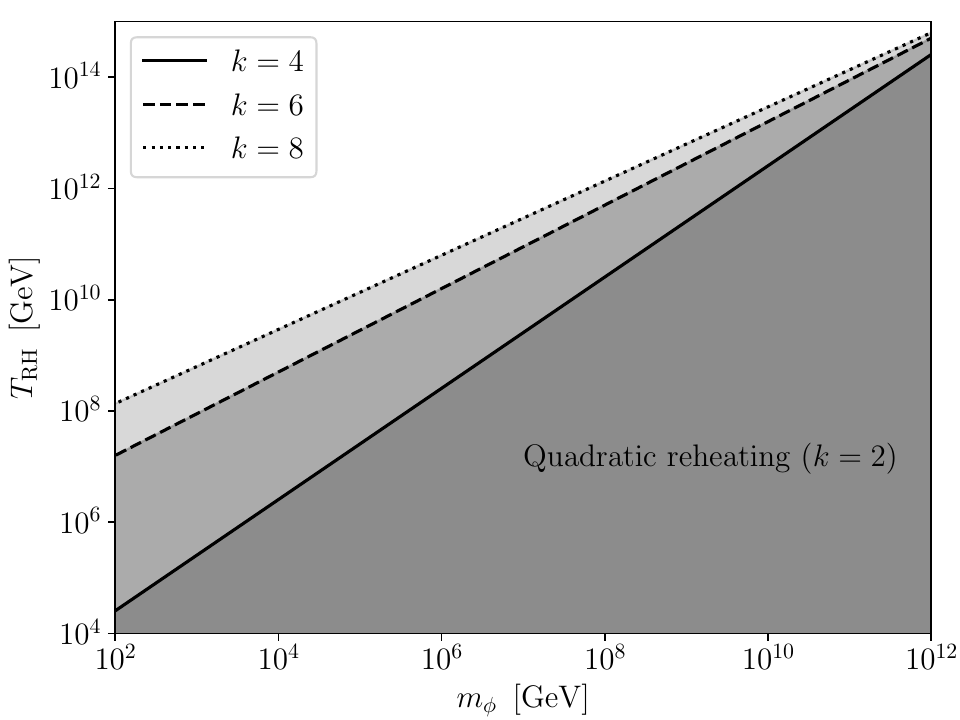}
\caption{\em \small Minimal reheating temperature from Eq.(\ref{Eq:limittrhgen}), below which the inflaton mass term drives the process, as a function of the bare mass $m_\phi$ and
for different values of $k$, $k=4$ (solid line), $k=6$ (dashed), $k=8$ (dotted). In the different shaded regions, reheating occurs while the inflaton oscillates in a quadratic potential $(k=2)$, given by its bare mass $m_\phi$. Above the lines, for different $k$, reheating occurs while the inflaton oscillates in the potential $V(\phi)\sim \phi^k$.}
\label{Fig:plot_num_sigma}
\end{figure}

\section{Consequence on the inflaton fragmentation}
\label{sec:frag}

Recently, the authors of \cite{GGMOPY} and \cite{Garcia:2023eol} have shown that fragmentation can significantly alter the reheating process. Indeed, the 
fragmentation of the inflaton condensate results in the 
population of an inflaton-particle bath, whose very low mass, 
proportional to the density of  the condensate which remains 
unfragmented, may not allow reheating temperatures above the BBN bounds for inflaton decays to fermions. This 
fragmentation is due to the presence of a self-scattering term 
of type $\lambda \phi^k$, with $k \geq 4$. The inflaton condensate does not fragment in the absence of self-interactions allowing for reheating to occur as discussed above. 

However, the study \cite{GGMOPY} was carried out in the context of 
a monomial potential of the type $V(\phi)=\lambda \phi^k$. 
It is then easy to see that the presence of a bare 
mass term of the type $\frac{1}{2}m_\phi^2\phi^2$ can 
change the conclusions of this study, in particular if the quadratic term begins to dominate {\it before} the 
fragmentation halts. If we define $\af$ as the value of the scale 
factor at the end of fragmentation, 
then
$\af/\aend=180,4.5\times10^4, 6\times 10^6$ and $7\times 10^8$, for $k=4,6,8$ and 10 respectively \cite{GGMOPY}.
In order for a quadratic term to affect the fragmentation process, we must have $\am \lesssim \af$
and using Eq.(\ref{Eq:am})
it becomes easy to compute, for each value of $k$, the minimal value of $m_\phi$ necessary to ensure
that the quadratic term dominates
the potential {\it before} the end of fragmentation. The problem of a leftover bath of massless inflatons can then be avoided
by stopping the fragmentation process. 

More precisely, when reheating begins, self interactions can source the growth of the inflaton fluctuations $\delta\phi(t,\boldsymbol{x})=\phi(t,\boldsymbol{x})-\bar{\phi}(t)$, where $\bar{\phi}$ denotes the homogeneous condensate. At early times, this growth can be captured by the linear equation of motion
\beq\label{eq:eomdeltaphi}
\ddot{\delta\phi} + 3H\dot{\delta\phi} - \frac{\nabla^2\delta\phi}{a^2} + V''(\bar{\phi})\,\delta\phi \;=\;0\,,
\eeq
where 
\beq\label{eq:vphiphi}
V''(\bar{\phi}) \;\simeq\; k(k-1) \lambda \bar{\phi}^{k-2} M_P^{4-k} + m_{\phi}^2\,.
\eeq
For $m_{\phi}=0$, the oscillating nature of this resulting effective mass term drives the resonant growth of $\delta\phi$ and the eventual fragmentation, $\delta\phi\gg \bar{\phi}$~\cite{Frolov:2010sz,Greene:1997fu,Kaiser:1997mp,Garcia-Bellido:2008ycs,Amin:2011hj,Hertzberg:2014iza,Lozanov:2016hid,Lozanov:2017hjm,Figueroa:2016wxr,Garcia:2023eol,GGMOPY}. However, if $m_{\phi}$ dominates before fragmentation, $V''\sim$~const., strongly suppressing the oscillatory driving force.\footnote{For a purely quadratic inflaton potential the growth of fluctuations is still present, albeit not exponentially enhanced, due to the coupling of $\delta\phi$ with the fluctuations of the metric~\cite{Finelli:1998bu,Jedamzik:2010dq,Martin:2019nuw}.}

\begin{figure}[!t]
\centering
\includegraphics[width=\columnwidth]{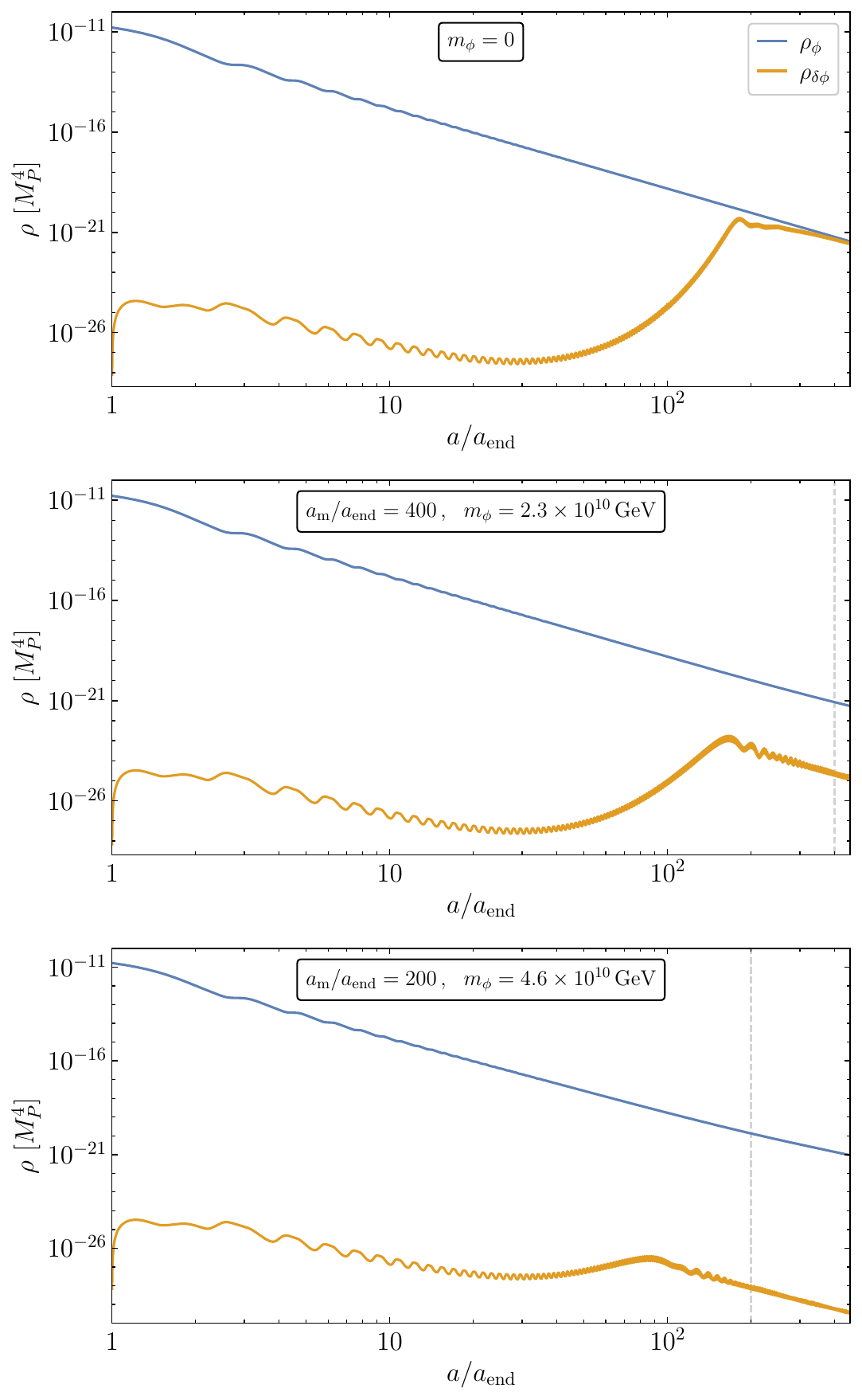}
\caption{\em \small Energy density of the inflaton fluctuations $\rho_{\delta\phi}$ compared to the total energy density $\rho_{\phi}$, for three values of the bare mass, for $k=4$. The vertical dashed line corresponds to the value of $\am/a_{\rm end}$ when $m_{\phi}\neq0$. In both of these cases, although $\am > a_{\rm F}$, the exponential growth of $\delta\phi$ is stopped by the transition to matter-domination.
}
\label{Fig:masss}
\end{figure}

Fig.~\ref{Fig:masss} shows the evolution of the total inflaton energy density $\rho_{\phi}$, compared to the energy density in its fluctuations $\rho_{\delta\phi}$, as computed numerically for a T-model of inflation \cite{Kallosh:2013hoa} with $k=4$ and three choices of the bare mass (see~\cite{Garcia:2023eol} for details). The top panel depicts the zero bare mass scenario. In it, the rapid growth of inflaton fluctuations driven by parametric resonance can be appreciated. This growth only stops when $\rho_{\phi}\simeq \rho_{\delta\phi}$ ($a/a_{\rm end}\simeq 180$), corresponding to the near-complete fragmentation of the inflaton condensate in favor of free $\phi$-particles.\footnote{The fragmentation of the inflaton condensate is not total even for $m_{\phi}=0$. A small but nonvanishing homogeneous component $\bar{\phi}$ remains, and its presence can induce the decay of the free inflaton quanta $\delta\phi$~\cite{GGMOPY,Garcia:2023eol}.} For the bottom two panels we take $m_{\phi}>0$. In both cases, the quartic $\rightarrow$ quadratic transition time has been chosen to be posterior to the complete fragmentation of the inflaton, $\am > a_{\rm F}$. A naive estimate from Eq.~(\ref{eq:vphiphi}) would indicate that the resonant growth of $\delta\phi$ would not stop until
\beq
\frac{a}{a_{\rm end}} \;=\; \frac{\sqrt{12\lambda}\phi_{\rm end}}{m_{\phi}} \;\simeq\; 2.6\,\frac{\am}{a_{\rm end}}\,,
\eeq
that is, the field would be fully fragmented before matter domination. However, the full numerical solution of the equation of motion (\ref{eq:eomdeltaphi}) shows that the growth of fluctuations is in reality suppressed from $a\lesssim \am/2$, as both panels of Fig.~\ref{Fig:masss} demonstrate. Therefore, reaching quadratic dominance is a sufficient condition to avert full fragmentation. Note that for smaller masses than those used in Fig.~\ref{Fig:masss}, fragmentation would nearly completely destroy the condensate and potentially disrupt the reheating process entirely.
On the other hand, for larger masses, the fragmentation process would not be operative at all. 

\begin{figure}[!ht]
\centering
\includegraphics[width=\columnwidth]{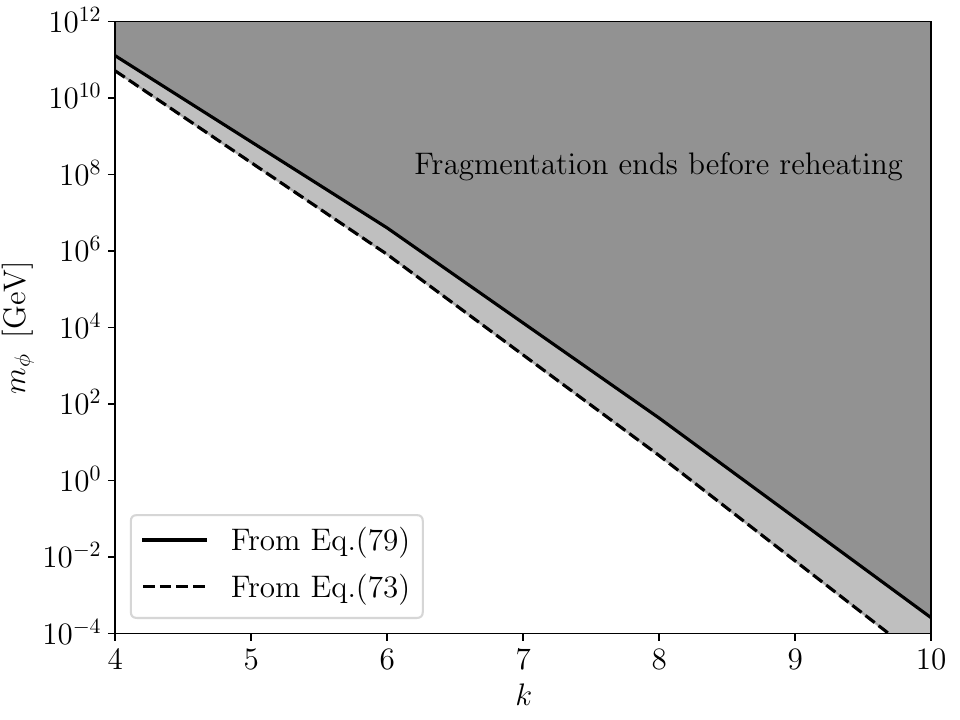}
\caption{\em \small Region in the parameter space where 
the fragmentation happens after the domination by the
bare mass term $\frac{1}{2}m_\phi^2 \phi^2$ over $\lambda \phi^k$, allowing for a quadratic reheating. The dashed line is obtained from Eq.(\ref{Eq:amk}), while the solid line is obtained from Eq.(\ref{eq:vphiphi}).
}
\label{Fig:plotfrag}
\end{figure}

A qualitative depiction of this result for potentials with $k \ge 4$ is shown in
Fig.~\ref{Fig:plotfrag}, where we plot the limit on the mass $m_\phi$ above which
the bare mass term dominates over $\lambda \phi^k$ in the potential as a function of $k$. We see that for larger 
value of $k$, where the fragmentation is less efficient due to the increasing difficulty for the self scattering to 
occurs for higher modes, even a small bare mass term can be sufficient to stop the fragmentation process 
and ensure a successful reheating.

\section{Conclusion}
\label{sec:sum}

Reheating in most models of inflation is accomplished through either inflaton decay or scattering to Standard Model particles. This typically after inflationary expansion ends and a period of inflaton oscillations begins.  When the potential is dominated by a quadratic term about its minimum, decays are necessary, as scatterings will not in general lead to a radiation dominated universe. However, 
potentials dominated by higher order interactions, $k > 2$, have anharmonic oscillations and scattering may lead to reheating, though these models may be subject to additional constraints arising from the fragmentation of the inflaton condensate.
In addition, the details of the reheating process and the final reheating temperature depend on the spin of the final state particles produced in the decay or scattering. 

In models of inflation for which the potential can be expanded about its minimum as $V(\phi) \propto \phi^k$, typically the lowest power, $k$ appearing in the expansion dominates the reheating process. For $k>2$, it is quite possible, as we have argued that in addition to the inflationary potential, a bare mass term in the full scalar potential is also present. This may arise from radiative corrections or supersymmetry breaking. 
In section \ref{masslim}, we derived upper limits to this mass from CMB observables. These limits are sufficiently weak so that the presence of the mass term may affect the reheating process.  Indeed, quite generally, if Eqs.~(\ref{Eq:limittrh}) for $k=4$ or (\ref{Eq:limittrhgen}) more generally are satified, the final reheating temperature will be determined by the quadratic rather than a higher order term. 

The qualitative effect of the mass term also depends on the reheating mechanism (decays or scattering) as well as the spin of the final states. For decays to fermions, the reheating temperature is increased by the presence of mass term, while for scalars, it is decreased. For decays to vectors, reheating does not occur for $k=4$ in the absence of a mass term and its presence allows for the possibility of reheating in this case. 
In contrast, if the mass term becomes important before the end of reheating for scattering to scalars, the reheating process is halted. Furthermore, when reheating is accomplished through scattering to scalars with $k \ge 4$, the density of inflatons is quickly redshifts (as $a^{-8}$) until the mass term comes to dominate. In this case, the residual inflaton matter density acts as cold dark matter and a strong limit on the inflaton mass has been derived in Eq.~(\ref{strlim}). Finally we have seen that for scattering to vectors, reheating with $k=4$ is not possible ($k \ge 6$ is required) and the mass term does not come to the rescue in this case.  

Understanding the reheating process after inflation is of great importance as it is not only responsible for providing an early period of radiation domination necessary for big bang nucleosynthesis, but may be the source of dark matter. 
Thermal production of dark matter in equilibrium remains an important mechanism, however, it is well established that non-equilibrium process just as freeze-in \cite{fimp} may also be the ultimate source of dark matter in the Universe. For these cases, a detailed understanding of reheating is essential and here we examined the role of a bare mass term for the inflaton in models where the inflationary dynamics are governed by higher order interactions. 

\vspace{0.5cm} 
\noindent
\acknowledgements
The authors thank Mathieu Gross and Jong-Hyun Yoon for extremely valuable discussions during the completion of our work. This project has received support from the European Union's Horizon 2020 research and innovation programme under the Marie Sklodowska-Curie grant agreement No 860881-HIDDeN, and the IN2P3 Master Projet UCMN. The work of M.A.G.G.~was supported by the DGAPA-PAPIIT grant IA103123 at UNAM, the CONAHCYT ``Ciencia de Frontera'' grant CF-2023-I-17, and the Programa de Investigaci\'on del Instituto de F\'isica 2023 (PIIF23). The work of K.A.O.~was supported in part by DOE grant DE-SC0011842 at the University of Minnesota.
The authors
acknowledge the support of the Institut Pascal at Universit\'e Paris-Saclay during the Paris-Saclay
Astroparticle Symposium 2023, with the support of the P2IO Laboratory of Excellence (program
“Investissements d’avenir” ANR-11-IDEX-0003-01 Paris-Saclay and ANR-10-LABX-0038) and the IN2P3 master project UCMN.

\section*{Appendix}

The decay (or scattering) rate of the inflaton, averaged over several oscillations, can be neatly 
expressed as \cite{GKMO2,gkkmov}
\beq
\Gamma_{\phi}(t) \;=\;
\gamma_{\phi}\left(\frac{\rho_{\phi}}{M_P^4}\right)^{l}\,,
\label{Eq:gammaphi}
\eeq
where
\beq
\label{Eq:gammaphibis}
\gamma_{\phi} \;=\; 
\begin{cases}
\sqrt{k(k-1)}\lambda^{1/k}M_P\dfrac{y_{\rm eff }^2}{8\pi}\,,\quad & \phi\rightarrow \bar{f}f\,,\\[10pt]
\dfrac{\mu_{\rm eff }^2}{8\pi\sqrt{k(k-1)}\lambda^{1/k}M_P}\,,\quad& \phi\rightarrow bb\,,\\[10pt]
\alpha^2 \left[k(k-1)\right]^\frac{3}{2}  \lambda^\frac{3}{k} M_P\,,\quad & \phi\rightarrow AA\,,\\[10pt] 
\dfrac{\sigma_{\rm eff}^2 M_P}{8\pi [k(k-1)]^{3/2}\lambda^{3/k}}\,,\quad & \phi\phi\rightarrow bb\,, \\[10pt] 
\beta^2 \left[k(k-1)\right]^\frac{1}{2} \lambda^\frac{1}{k} M_P\,,\quad & \phi\phi\rightarrow AA\,, 
\end{cases}
\eeq
and
\beq
l \;=\; 
\begin{cases}
\frac{1}{2}-\frac{1}{k}\,,\quad & \phi\rightarrow \bar{f}f\,,\\
\frac{1}{k}-\frac{1}{2}\,,\quad & \phi\rightarrow bb\,,\\
\frac32-\frac3k\,,\quad & \phi\rightarrow AA\,,\\
\frac{3}{k}-\frac{1}{2}\,,\quad & \phi\phi\rightarrow bb
\,,\\
\frac{3}{2}-\frac{1}{k}\,,\quad & \phi\phi\rightarrow AA\,.
\end{cases}
\label{ls}
\eeq

So long as $\gamma_\phi \ll H$, the Friedmann equation for $\rho_\phi$ (\ref{Eq:frphi}) can be integrated to give
\beq
\rho_\phi(a) = \rho_{\rm end} \left(\frac{a}{a_{\rm end}} \right)^{-\frac{6k}{k+2}} \, ,
\label{Eq:rhophi}
\eeq
which sources the Boltzmann equation (\ref{Eq:radiation_density}). 
This can be rewritten as
\beq
\frac{1}{a^4} \frac{d }{da}\left( \rho_R a^4\right) \;=\; \frac{2k}{k+2}
\frac{\gamma_\phi}{aH}
\frac{\rho_\phi^{l+1}}{M_P^{4l}} \, ,
\label{diffa}
\eeq
and can be integrated to give
\beq
\rho_R \; = \; \frac{2 k}{k + 8 - 6kl} \frac{\gamma_\phi}{H_{\rm end}}
 \frac{\rho_{\rm end}^{l+1}}{M_P^{4l}} \left( \frac{a_{\rm end}}{a} \right)^4
\left[ \left( \frac{a}{a_{\rm end}}\right)^{\frac{k+8-6kl}{k+2}} -1\right] \, ,
\label{Eq:rhoR}
\eeq
where $H^2_{\rm end} = \rho_{\rm end}/3 M_P^2$. 
At later times when $a \gg a_{\rm end}$ and $8+k-6kl>0$,
we can approximate $\rho_R$ as
\beq
\rho_R^{a\gg a_{\rm end}} \; = \; \frac{2 k}{k + 8 - 6kl} \frac{\gamma_\phi}{H_{\rm end}}
 \frac{\rho_{\rm end}^{l+1}}{M_P^{4l}} \left( \frac{a_{\rm end}}{a} \right)^{\frac{3k + 6kl}{k+2}} \, .
\label{Eq:rhoRapp}
\eeq
If  $8+k-6kl<0$,
\beq
\rho_R^{a\gg a_{\rm end}} \;=\; \frac{2k}{6kl-k-8} \frac{\gamma_\phi}{H_{\rm end}} \frac{\rho_{\rm end}^{l+1}}{M_P^{4l}}  \left(\frac{a_{\rm end}}{a}\right)^{4}\,,
\label{Eq:rhoRappter}
\eeq
which implies that the temperature would simply redshift as $T\propto a^{-1}$.

Finally, when $\rho_R(T_{\rm RH}) = \rho_\phi(T_{\rm RH})$,
we obtain \cite{GKMO2}
\beq
\frac{a_{\rm RH}}{a_{\rm end}}=\left[ \frac{k+8-6kl}{2k} \frac{M_P^{4l-1}\rho_{\rm end}^{\frac{1}{2}-l}}{\sqrt{3}\gamma_\phi}\right]^{\frac{k+2}{3k-6kl}} \, ,
\eeq
for $8+k-6kl >0$ and for $8+k-6kl < 0$, 
\beq
\frac{a_{\rm RH}}{a_{\rm end}}=\left[ \frac{6kl-k-8}{2k} \frac{M_P^{4l-1}\rho_{\rm end}^{\frac{1}{2}-l}}{\sqrt{3} \gamma_\phi}\right]^{\frac{k+2}{2k-8}} \, .
\label{arhaendneg}
\eeq
Note that Eq.~(\ref{arhaendneg}) is only true for $k>4$.
When $k \le 4$ and $8+k-6kl < 0$, reheating never occurs.

Evaluating $\rho_R$ at $a = \arh$ gives
\begin{align}
T_{\rm RH} \;&=\; \left(\frac{1}{\alpha} \right)^{\frac{1}{4}}
\left[\frac{2k}{k+8-6kl} 
\frac{\sqrt{3}\gamma_\phi}{M_P^{4l-1}} \right]^{\frac{1}{2-4l}} \, ,
\label{trhpos}
\end{align}
for $8+k-6kl > 0$, and 
\begin{align}
T_{\rm RH} \;&=\; \left(\frac{1}{\alpha} \right)^{\frac{1}{4}}
\left[\frac{2k}{6kl-k-8} 
\frac{\sqrt{3}\gamma_\phi}{M_P^{4l-1}} \rho_{\rm end}^{\frac{6kl-k-8}{6k}}\right]^{\frac{3k}{4k-16}} \, .
\label{trhneg}
\end{align}
for $8+k-6kl < 0$ and $k>4$

\end{document}